\documentstyle[aps,amstex,amssymb,graphicx,preprint]{revtex}
\begin{document}
%
%%%%%%%%%%%%%%%%%%%%%%%%%%%%%%%%%%%%%%%%%%%%%%%%%%%%%%%%%%%%%%%%%%%%%%%
%   TITLE
%%%%%%%%%%%%%%%%%%%%%%%%%%%%%%%%%%%%%%%%%%%%%%%%%%%%%%%%%%%%%%%%%%%%%%%
%

\title{Ab--initio calculation of Kerr spectra for semi--infinite
systems including multiple reflections and optical interferences }
\author{A. Vernes}
\address{Center for Computational Materials Science, \\
Technical University Vienna, Gumpendorferstr. 1a, A--1060 Vienna, Austria} 
\author{L. Szunyogh}
\address{Center for Computational Materials Science, \\
Technical University Vienna, Gumpendorferstr. 1a, A--1060 Vienna, Austria\\
Department of Theoretical Physics, Budapest University of Technology and
Economics, \\
Budafoki \'{u}t 8, H--1521 Budapest, Hungary}
\author{P. Weinberger}
\address{Center for Computational Materials Science, \\
Technical University Vienna, Gumpendorferstr. 1a, A--1060 Vienna, Austria}
%% \date{\today}
\date{November 27, 2001}
\maketitle
%
%%%%%%%%%%%%%%%%%%%%%%%%%%%%%%%%%%%%%%%%%%%%%%%%%%%%%%%%%%%%%%%%%%%%%%%%%%%
%                                STATUS OF PAPER
%%%%%%%%%%%%%%%%%%%%%%%%%%%%%%%%%%%%%%%%%%%%%%%%%%%%%%%%%%%%%%%%%%%%%%%%%%%
%
%% \vspace{-90ex}
%% \begin{flushright}
%% \textbf{\large submitted to PRB}\\
%% \end{flushright}
%% \vspace{80ex}
%
%
%%%%%%%%%%%%%%%%%%%%%%%%%%%%%%%%%%%%%%%%%%%%%%%%%%%%%%%%%%%%%%%%%%%%%%%
%   ABSTRACT
%%%%%%%%%%%%%%%%%%%%%%%%%%%%%%%%%%%%%%%%%%%%%%%%%%%%%%%%%%%%%%%%%%%%%%%
%

\begin{abstract}
  Based on Luttinger's formulation the complex optical conductivity
  tensor is calculated within the framework of the spin--polarized
  relativistic screened Korringa--Kohn--Rostoker method for layered
  systems by means of a contour integration technique.  For polar
  geometry and normal incidence ab--initio Kerr spectra of multilayer
  systems are then obtained by including via a $2\times 2$ matrix
  technique all multiple reflections between layers and optical
  interferences in the layers. Applications to Co$\mid$Pt$_5$ and
  Pt$_3\mid$Co$\mid$Pt$_5$ on the top of a semi--infinite fcc--Pt(111)
  bulk substrate show good qualitative agreement with the experimental
  spectra, but differ from those obtained by applying the commonly
  used two--media approach.
\end{abstract}

\draft

%%\preprint{IP/1}

%%\narrowtext
%
%%%%%%%%%%%%%%%%%%%%%%%%%%%%%%%%%%%%%%%%%%%%%%%%%%%%%%%%%%%%%%%%%%%%%%%
%   INTRODUCTION
%%%%%%%%%%%%%%%%%%%%%%%%%%%%%%%%%%%%%%%%%%%%%%%%%%%%%%%%%%%%%%%%%%%%%%%
%

\section{Introduction}

\label{sect:intro}

Magneto-optical effects not only provide a powerful tool in probing
the magnetic properties of solids, \cite{BSE90,HTB97,SKB+98} but are
also of direct technological interest as phenomena to be used for
high--density magneto--optical recording. \cite{HTB97,BS94} Up to now,
however, realistic theoretical investigations were lacking, because
band--structure methods using supercells, \cite{OMSK92} cannot provide
an adequate description of layered systems, for which special
computational techniques such as the spin--polarized relativistic
screened Korringa--Kohn--Rostoker (SKKR) method have been designed.
\cite{SUWK94,SUW95,USW95} Furthermore, the absorptive parts of the
optical conductivity tensor as obtained from the inter--band
contributions, \cite{Cal74} are not sufficient for magneto--optical
Kerr spectra calculations, since also the dissipative parts have to be
known. Hence, in supercell type calculations, besides the necessity to
use the Kramers--Kronig relations, one has to include also the
inter--band contributions by means of a semi--empirical Drude term.
\cite{Opp99} Only recently a better scheme was developed by two of the
present authors \cite{SW99}, in which a contour integration is used to
obtain the complex optical conductivity tensor as based Luttinger's
formula, \cite{Lut67} which in turn includes all inter-- and
intra--band contributions. \cite{VSW01b} Combining this contour
integration technique with the SKKR method, realistic inter-- and
intra--layer complex optical conductivities can be obtained for
layered systems.

Having evaluated the inter-- and intra--layer optical conductivities,
the magneto--optical Kerr spectra can then be calculated by using a
macroscopical model such as, e.g., the two--media approach.
\cite{RS90} Because the layered system contains more boundaries than
just the interface between the vacuum and the surface layer, the
two--media approach not fully includes the dynamics of the
electromagnetic waves propagation in such systems.  Since the
pioneering work of Abel\'{e}s in 1950, \cite{Abe50} several methods
are known in the literature \cite {Smi65,Hun67} to treat multiple
reflections and interferences using either a $2\times 2$ matrix
\cite{Man90,Man95} or $4\times 4$ matrix \cite{Yeh80,ZML+90,AL93}
technique.  In the present paper the magneto--optical Kerr spectra of
layered systems are evaluated for the most frequently used
experimental set--up, namely polar geometry and normal incidence, by
making use of the complex optical conductivity tensor and the $2\times
2$ matrix technique.

In Section \ref{sect:theory} the theoretical background is reviewed
briefly. Computational aspects are then summarized in Section
\ref{sect:comput}.  In Section \ref{sect:moke} the two--media approach
(Sect.\ \ref{sect:twomed}) and the applied $2\times 2$ matrix
technique (Sect.\ \ref{sect:2x2}) are viewed as two different
macroscopic models of how to calculate magneto--optical Kerr spectra
of layered systems. Particular emphasis has been put in Sect.\
\ref{sect:lreps} on the construction of layer--resolved permittivities
in terms of the (macroscopic) material equation within linear
response. This construction method combined with the $2\times 2$
matrix technique, allows one to determine self--consistently
layer--resolved permittivities, see Sect.\ \ref{sect:sclreps}. As an
illustration ab--initio Kerr spectra of Co$\mid $Pt multilayer systems
are presented and discussed in Section \ref{sect:res}.  Finally, in
Section \ref{sect:summary} the main results are summarized.

%
%%%%%%%%%%%%%%%%%%%%%%%%%%%%%%%%%%%%%%%%%%%%%%%%%%%%%%%%%%%%%%%%%%%%%%%
%   THEORETICAL FRAMEWORK
%%%%%%%%%%%%%%%%%%%%%%%%%%%%%%%%%%%%%%%%%%%%%%%%%%%%%%%%%%%%%%%%%%%%%%%
%

\section{Theoretical framework}

\label{sect:theory}

%
%%%%%%%%%%%%%%%%%%%%%%%%%%%%%%%%%%%%%%%%%%%%%%%%%%%%%%%%%%%%%%%%%%%%%%%
%   LUTTINGER FORMALISM
%%%%%%%%%%%%%%%%%%%%%%%%%%%%%%%%%%%%%%%%%%%%%%%%%%%%%%%%%%%%%%%%%%%%%%%
%

\subsection{Luttinger's formalism}

\label{sect:luttinger}

The frequency dependent complex optical conductivity tensor $\tilde{
  \bbox{\sigma}}(\omega )$ can be evaluated starting from the
well--known Kubo formula and using a scalar potential description of
the electric field.  \cite{Kub57} However, by using the equivalent
\cite{Hu93} vector potential description of the electric field, one
ends up with Luttinger's formula: \cite{Lut67}
\begin{equation}
\tilde{\sigma}_{\mu \nu }(\omega )=%
{\displaystyle{\tilde{\Sigma}_{\mu \nu }(\omega )-\tilde{\Sigma}_{\mu
\nu }(0) \overwithdelims.. \hbar \omega +i\delta }}%
\ ,  \label{eq:cplxsigma}
\end{equation}
with the current--current correlation function as given by \cite{Mah90} 
\begin{equation}
\tilde{\Sigma}_{\mu \nu }(\omega )=\frac{i\hbar }{V}\,\sum_{m,n}\,%
{f(\epsilon _{n})-f(\epsilon _{m}) \overwithdelims.. \hbar \omega +
i\delta +(\epsilon _{n}-\epsilon _{m})}%
J_{nm}^{\mu }J_{mn}^{\nu }\ .  \label{eq:curcurcorfct}
\end{equation}
Here $f(\epsilon )$ is the Fermi--Dirac distribution function,
$\epsilon _{m},\epsilon _{n}$ a pair of eigenvalues of the
one--electron Hamiltonian, the $J_{mn}^{\mu }$ are matrix elements of
the electronic current operator ($\mu = {\rm x,y,z}$) and $V$ the
reference (crystalline) volume.

The positive infinitesimal $\delta $ implies the electromagnetic field
to be turned on at $t=-\infty $ and hence describes the interaction of
the system with its surrounding. \cite{Lax58} However, as can be seen
from Eq.\ (\ref{eq:curcurcorfct}), $\delta $ can be also viewed as a
finite life--time broadening, which accounts for all scattering
processes at a finite temperature.

Luttinger's formula (\ref{eq:cplxsigma}) and Eq.\
(\ref{eq:curcurcorfct}) have several advantages over the commonly used
optical conductivity tensor formula of Callaway. \cite{Cal74} First of
all, in contrast to the latter, Eq.\ (\ref{eq:cplxsigma})
simultaneously provides both, the absorptive and the dissipative parts
of the optical conductivity tensor. Hence there is no need for using
the Kramers--Kronig relations in the Luttinger's formalism. On the
other hand, as recently was shown \cite{VSW01b} Luttinger's formalism
accounts for both, the inter-- and the intra--band contribution on the
same footing. Thus by using Eq.\ (\ref {eq:cplxsigma}) in combination
with Eq.\ (\ref{eq:curcurcorfct}), one does not need to include a
phenomenological Drude term in order to simulate the intra--band
contribution. \cite{OMSK92} Furthermore, as was also demonstrated
\cite{VSW01b} Eqs.\ (\ref{eq:cplxsigma}) and (\ref{eq:curcurcorfct})
can be used for calculations in the static ($\omega =0$) limit,
provided the life--time broadening is kept finite ($\delta \neq 0$).

%
%%%%%%%%%%%%%%%%%%%%%%%%%%%%%%%%%%%%%%%%%%%%%%%%%%%%%%%%%%%%%%%%%%%%%%%
%   CONTOUR INTEGRATION TECHNIQUE
%%%%%%%%%%%%%%%%%%%%%%%%%%%%%%%%%%%%%%%%%%%%%%%%%%%%%%%%%%%%%%%%%%%%%%%
%

\subsection{Contour integration technique}

\label{sect:contour}

Instead of evaluating the sums in the expression for the
current--current correlation function (\ref{eq:curcurcorfct}) over
eigenvalues of the one--electron Hamiltonian, $\tilde{\Sigma}_{\mu \nu
  }(\omega )$ can be calculated by means of a contour integration in
the complex energy plane.

For the selection of a particular contour $\Gamma $, this technique
\cite{SW99} exploits the facts that with the exception of the
Matsubara poles $z_{k}=\varepsilon _{{\rm F}}+i\left( 2k-1\right)
\delta _{T}$ ($k=0,\pm 1,\pm 2,\ldots\ $, and $\delta _{T}=\pi k_{{\rm
B}}T$), \cite{NSW+94} in both semi--planes the Fermi--Dirac
distribution function of complex argument $f(z)$ is analytical
\cite{Mah90} and is a real function for complex energies $
z=\varepsilon \pm i\delta _{j}$ situated in--between two successive
Matsubara poles. \cite{WLZ+95} The latter property of $f(z)$, e.g., is
exploited by using $\delta _{j}=2N_{j}\delta _{T} $, where $N_{1}$ is
the number of Matsubara poles included in $\Gamma $ in the upper
semi--plane and $N_{2}$ in the lower semi--plane. \cite{SW99}

By applying the residue theorem, it has been shown \cite{SW99} that
equivalently to Eq.\ (\ref{eq:curcurcorfct}) one has
\begin{align}
\tilde{\Sigma}_{\mu \nu }(\omega )& =\int_{\Gamma }\hspace{-3.0ex}%
\circlearrowright \ dz\ f(z)\ \tilde{\Sigma}_{\mu \nu }(z+\zeta ,z)-\left[
\int_{\Gamma }\hspace{-3.0ex}\circlearrowright \ dz\ f(z)\ \tilde{\Sigma}%
_{\mu \nu }(z-\zeta ^{\ast },z)\right] ^{\ast }  \nonumber \\
& -2i\delta _{T}\sum_{k=-N_{2}+1}^{N_{1}}\left[ \tilde{\Sigma}_{\mu \nu
}(z_{k}+\zeta ,z_{k})+\tilde{\Sigma}_{\mu \nu }^{\ast }(z_{k}-\zeta ^{\ast
},z_{k})\right] \ ,  \label{eq:Sgmw}
\end{align}
such that

\begin{equation}
\tilde{\Sigma}_{\mu \nu }(0)=\int_{\Gamma }\hspace{-3.0ex}\circlearrowright
\ dz\ f(z)\ \tilde{\Sigma}_{\mu \nu }(z,z)-2i\delta
_{T}\sum_{k=-N_{2}+1}^{N_{1}}\tilde{\Sigma}_{\mu \nu }(z_{k},z_{k})\ ,
\label{eq:Sgm0}
\end{equation}
where $\zeta =\hbar \omega +i\delta $ and the kernel 
\begin{equation}
\tilde{\Sigma}_{\mu \nu }(z_{1},z_{2})=-\frac{\hbar }{2\pi V}\ {\rm Tr\ }%
\left[ J^{\mu }\ G(z_{1})\ J^{\nu }\ G(z_{2})\right] \ ,  \label{eq:Sgmzz}
\end{equation}
is related to the electronic Green function $G(z)$. The auxiliary
quantity $\tilde{\Sigma}_{\mu \nu }(z_{1},z_{2})$ has already been
used in residual resistivity calculations ($\omega ,T=0$) of
substitutionally disordered bulk systems \cite{But85} and
magneto--transport calculations of inhomogeneously disordered layered
systems. \cite{WLB+96} Only recently, however, it has been shown,
\cite{VSW01b} that Eqs.\ (\ref{eq:Sgmw})--(\ref{eq:Sgmzz}) preserve
all the advantages and features of Luttinger's formalism as was
mentioned already above.

In the present paper, $\tilde{\Sigma}_{\mu \nu }(z_{1},z_{2})$ is
evaluated in terms of relativistic current operators \cite{WLB+96} and
the Green functions provided by the spin--polarized relativistic SKKR
method for layered systems \cite{SUWK94,SUW95,USW95}. The optical
conductivity tensor of a multilayer system is then given \cite{VSW01c}
by
\begin{equation}
\tilde{\bbox{\sigma}}(\omega )=\sum_{p=1}^{N}\sum_{q=1}^{N}\tilde{%
\bbox{\sigma}}^{pq}(\omega )\ ,  \label{eq:totcmpxoptcond}
\end{equation}
with $\tilde{\bbox{\sigma}}^{pq}(\omega )$ refering to either the
inter-- ($p\neq q$) or the intra--layer ($p=q$) contribution to the
optical conductivity tensor.

%
%%%%%%%%%%%%%%%%%%%%%%%%%%%%%%%%%%%%%%%%%%%%%%%%%%%%%%%%%%%%%%%%%%%%%%%
%   COMPUTATIONAL DETAILS
%%%%%%%%%%%%%%%%%%%%%%%%%%%%%%%%%%%%%%%%%%%%%%%%%%%%%%%%%%%%%%%%%%%%%%%
%

\section{Computational details}

\label{sect:comput}

In addition to the number of Matsubara poles considered, the optical
conductivity tensor depends also on the number of complex energy
points $n_{z}$ used in order to evaluate the energy integrals in Eqs.\ 
(\ref{eq:Sgmw}) and (\ref{eq:Sgm0}), on the number of $\Vec{k}$
--points used to calculate the scattering path operators that define
the Green functions \cite{SUW95} and $\tilde{ \Sigma} _{\mu \nu }(z\pm
\hbar \omega +i\delta ,z)$ for a given energy $z$.  Recently, the
present authors have proposed two schemes to control the accuracy of
these $z$-- and $\Vec{k}$--integrations. \cite{VSW01a}

The first of these schemes is meant to control the accuracy of the
$z$--integrations along each contour part by comparing the results
obtained from the Konrod quadrature \cite{Lau97,CGG+00}, ${\cal
K}_{2n_{z}+1} \tilde{\Sigma}_{\mu \nu }(\omega )$, with those from the
Gauss integration rule, ${\cal G}_{n_{z}}\tilde{ \Sigma}_{\mu \nu
}(\omega )$. \cite{PFT+92} On a particular part of the contour
$\tilde{\Sigma} _{\mu \nu }(\omega )$ is said to be converged, if the
following convergence criterion: \cite{VSW01a}
\begin{equation}
\max \ \left| \ {\cal K}_{2n_{z}+1}\tilde{\Sigma}_{\mu \nu }(\omega )- {\cal %
G}_{n_{z}}\tilde{\Sigma}_{\mu \nu }(\omega )\ \right| \leq \varepsilon _{z}\,,
\label{eq:zconv}
\end{equation}
is fulfilled for a given accuracy parameter $\varepsilon _{z}$.

The other scheme refers to the cumulative special points method,
\cite{VSW01a} which permits to perform two--dimensional $\vec{k}$
--space integrations with arbitrary high precision. This method
exploits the arbitrariness of the special points mesh origin.
\cite{Haw92} For a given (arbitrary high) accuracy $\epsilon
_{\Vec{k}}$ the following convergence criterion has to apply
\begin{equation}
\max \ \left| \ {\cal S}_{n_{i}}\tilde{\Sigma}_{\mu \nu }(z^{\prime },z)-%
{\cal S}_{n_{i-1}}\tilde{\Sigma}_{\mu \nu }(z^{\prime },z)\ \right| \leq
\varepsilon _{\vec{k}}\,,  \label{eq:kconv}
\end{equation}
for any complex energy $z$ on the contour or $z_{k}$ Matsubara
pole. Here $n_{i}=2^{i+2}n_{0}$ ($n_{0}\in N$) is the number of
divisions along each primitive translation vector in the
two--dimensional $\Vec{k}$--space and $z^{\prime }=z+\zeta$, $z-\zeta
^{\ast }$.

In the present paper, the optical conductivity tensor calculations
were carried out for $T=300$ K, using a life--time broadening of
$0.048$ Ryd and\ $N_{2}=2$ Matsubara poles in the lower
semi--plane. Because the computation of $\tilde{\sigma}_{\mu \nu
}(\omega )$ does not depend on the form of the contour, \cite{VSW01a}
in the upper semi--plane we have accelerated the calculations by
considering $N_{1}=37$ Matsubara poles. The convergence criteria (\ref
{eq:zconv}) and (\ref{eq:kconv}) were fulfilled for $\varepsilon
_{z}=\varepsilon _{\Vec{k}}=10^{-3}$ a.u. .

%
%%%%%%%%%%%%%%%%%%%%%%%%%%%%%%%%%%%%%%%%%%%%%%%%%%%%%%%%%%%%%%%%%%%%%%%
%   MAGNETO-OPTICAL KERR EFFECT
%%%%%%%%%%%%%%%%%%%%%%%%%%%%%%%%%%%%%%%%%%%%%%%%%%%%%%%%%%%%%%%%%%%%%%%
%

\section{Magneto--optical Kerr effect}

\label{sect:moke}

In the case of the polar magneto--optical Kerr effect (PMOKE), \cite{RS90}
the Kerr rotation angle
\begin{equation}
\theta _{{\rm K}}=-\ \frac{1}{2}\left( \Delta _{+}-\Delta _{-}\right)
\label{eq:exactkang}
\end{equation}
and the Kerr ellipticity
\begin{equation}
\varepsilon _{{\rm K}}=-\ \frac{r_{+}-r_{-}}{r_{+}+r_{-}}\ 
\label{eq:exactkell}
\end{equation}
are given in terms of the complex reflectivity of the right-- (+) and
left--handed (--) circularly polarized light
\begin{equation}
\tilde{r}_{\pm }=\frac{{\cal E}_{\pm }^{({\rm r})}}{{\cal E}^{({\rm i})}}%
=r_{\pm }\ e^{i\Delta _{\pm }}\ .  \label{eq:cmplxref}
\end{equation}
Here the complex amplitude of the reflected right-- and left--handed
circularly polarized light is denoted by ${\cal E}_{\pm }^{({\rm r})}$
and that of the incident light by ${\cal E}^{({\rm i})}$; $\Delta
_{\pm }$ is the phase of the complex reflectivity $\tilde{r}_{\pm }$\ 
and $r_{\pm }=\left| \tilde{r}_{\pm }\right| $. Eqs.\ 
(\ref{eq:exactkang}) and (\ref {eq:exactkell}) are exact, which can be
easily deduced from simple geometrical arguments. However, in order to
apply these relations, one needs to make use of a macroscopic model
for the occurring reflectivities.

%
%%%%%%%%%%%%%%%%%%%%%%%%%%%%%%%%%%%%%%%%%%%%%%%%%%%%%%%%%%%%%%%%%%%%%%%
%   TWO-MEDIA APPROACH
%%%%%%%%%%%%%%%%%%%%%%%%%%%%%%%%%%%%%%%%%%%%%%%%%%%%%%%%%%%%%%%%%%%%%%%
%

\subsection{Macroscopic model I: the two--media approach}

\label{sect:twomed}

This simplest and most commonly used macroscopic model treats the
multilayer system as a homogeneous, anisotropic, semi--infinite
medium, such that the incident light is reflected only at the boundary
between the vacuum and the surface (top) layer.

In case of normal incidence the two--media approach provides an
appropriate formula for the complex Kerr angle \cite{RS90}
\begin{equation}
\tilde{\Phi}_{{\rm K}}\equiv \theta _{{\rm K}}-i\varepsilon _{{\rm K}}=i%
\frac{\tilde{r}_{+}-\tilde{r}_{-}}{\tilde{r}_{+}+\tilde{r}_{-}}
\ ,  \label{eq:exactcmplxkangni}
\end{equation}
which can be deduced from Eqs.\ (\ref{eq:exactkang}) and
(\ref{eq:exactkell}) by assuming a small difference in the complex
reflectivity of the right-- and left--handed circularly polarized
light. Because ${\rm \mathop{\rm Im}\ }\tilde{\sigma}_{{\rm xy}}\left(
  \omega \right) $ usually is almost a hundred times smaller than
${\rm Re\ }\tilde{\sigma}_{{\rm xx}}\left( \omega \right) $, see Ref.\ 
\cite{VSW01b}, the average complex refractive index of the
right-- and left--handed circularly polarized light is dominated by $%
\tilde{\sigma}_{{\rm xx}}\left( \omega \right) $ and hence the
following direct formula results from Eq.\
(\ref{eq:exactcmplxkangni}): \, \cite{RS90}
\begin{equation}
\tilde{\Phi}_{{\rm K}}\approx \frac{\tilde{\sigma}_{{\rm xy}}\left( \omega
\right) }{\tilde{\sigma}_{{\rm xx}}\left( \omega \right) }\frac{1}{\sqrt{1-\ 
{\displaystyle{4\pi i \overwithdelims.. \omega }}%
\tilde{\sigma}_{{\rm xx}}\left( \omega \right) }}\ ,  \label{eq:cmplxkangni}
\end{equation}
with $\tilde{\bbox{\sigma}}(\omega )$ as given by Eq.\
(\ref{eq:totcmpxoptcond}). It should be noted that the ``direct''
formula in Eq.\ (\ref{eq:cmplxkangni}) was introduced by Reim and
Schoenes \cite{RS90} in order to extract the optical conductivity
tensor elements $\tilde{\sigma}_{{\rm xx}}\left( \omega \right) $ and
$\tilde{\sigma}_{{\rm xy}}\left( \omega \right) $ from experimental
PMOKE data.

%
%%%%%%%%%%%%%%%%%%%%%%%%%%%%%%%%%%%%%%%%%%%%%%%%%%%%%%%%%%%%%%%%%%%%%%%
%   LAYER-RESOLVED PERMITTIVITY
%%%%%%%%%%%%%%%%%%%%%%%%%%%%%%%%%%%%%%%%%%%%%%%%%%%%%%%%%%%%%%%%%%%%%%%
%

\subsection{Macroscopic model II: layer--resolved permittivities}

\label{sect:lreps}

Within linear response theory \cite{AM76} the Fourier transformed
macroscopic material equations, \cite{AG66} averaged over the
reference volume $V$, directly yields the total electric displacement
\begin{equation}
\frac{1}{V}\int_{V}d^{3}r\ \vec{D}\left( \vec{r},\omega \right) =\frac{1}{V}%
\int_{V}d^{3}r\int_{V}d^{3}r^{\prime }\ \tilde{\bbox{\epsilon}}\left( \omega
;\vec{r},\vec{r}^{\prime }\right) \vec{E}\left( \vec{r}^{\prime },\omega
\right) \ ,  \label{eq:intmateqrw}
\end{equation}
provided that the dielectric function $\tilde{\bbox{\epsilon}}\left(
\omega ;\vec{r},\vec{r}^{\prime }\right) $ and the Fourier components
of the electric field $\vec{E}\left( \vec{r}^{\prime },\omega \right)
$ are known.

Using non--overlapping cells in configuration space (Atomic Sphere
Approximation (ASA), applied in the present approach), the reference
volume can be written as
\[
V=\ \sum_{p=1}^{N}\left( N_{\parallel }\sum_{i}\Omega _{pi}\right) \equiv
\sum_{p=1}^{N}\Omega ^{p}\ , 
\]
where $N_{\parallel }$ is the number of atoms per layer (the same 2D
lattice has to apply for each layer $p$), $N$ the total number of
layers and $\Omega _{pi}$ the volume of the $i$th atomic sphere in
layer $p$.

Assuming that plane waves propagate in a layer like in a
two--dimensional unbound homogeneous medium and that $\vec{D
  }_{pi}\left( \vec{r},\omega \right) =\vec{D}_{p}\left(
  \vec{r},\omega \right) $, the integral on the left hand side of Eq.\ 
(\ref{eq:intmateqrw}) can be written within the ASA as
\begin{equation}
\int_{V}d^{3}r\ \vec{D}\left( \vec{r},\omega \right) =N_{\parallel
}\sum_{p=1}^{N}{\cal \vec{D}}_{p}\sum_{i}\Omega _{pi}\left[
1+6\sum_{k=1}^{\infty }\frac{\left( -1\right) ^{k}\left( k+1\right) }{\left(
2k+3\right) !}\left( \frac{2\pi }{\lambda _{0}}n_{p}S_{pi}\right) ^{2k}%
\right] \ ,  \label{eq:lhsmateq}
\end{equation}
where ${\cal \vec{D}}_{p}$ is the amplitude of the electric
displacement, $\vec{n}_{p}$ is the refraction vector,
\begin{equation}
\vec{n}_{p}=\frac{\vec{q}_{p}}{q_{0}}\ ,  \label{eq:defrefvec}
\end{equation}
$\vec{q}_{p}$ the wave vector ($q_{0}=2\pi /\lambda _{0}$ refers to
the propagation constant in vacuum) and $S_{pi}$ is the radius of the
$i$th atomic sphere in layer $p$. Accordingly, the double integral on
the right hand side of Eq.\ (\ref{eq:intmateqrw}) reduces to
\begin{eqnarray}
\int_{V}d^{3}r\int_{V}d^{3}r^{\prime }\ \tilde{\bbox{\epsilon}}\left( \omega
;\vec{r},\vec{r}^{\prime }\right) \vec{E}\left( \vec{r}^{\prime },\omega
\right) &=&\left( 4\pi \right) ^{2}N_{\parallel }\sum_{p,q=1}^{N}{\cal \vec{E%
}}_{q}\sum_{i,j}\int_{0}^{S_{pi}}dr\ r^{2}\int_{0}^{S_{qj}}dr^{\prime }\
\left( r^{\prime }\right) ^{2}  \nonumber \\
&&\tilde{\bbox{\epsilon}}^{pi,qj}\left( \omega ;r,r^{\prime }\right) \ \left[
1+\sum_{k=1}^{\infty }\frac{\left( -1\right) ^{k}}{\left( 2k+1\right) !}%
\left( \frac{2\pi }{\lambda _{0}}n_{q}r^{\prime }\right) ^{2k}\right] \ ,
\label{eq:rhsmateq}
\end{eqnarray}
where ${\cal \vec{E}}_{q}$ is the amplitude of the electric field in
layer $q$ and $\bbox{\tilde{\epsilon}}^{pi,qj}\left( \omega
  ;r,r^{\prime }\right) $ is the dielectric function
$\tilde{\bbox{\epsilon}}\left( \omega ;\vec{r},\vec{r }^{\prime
    }\right) $ at $\vec{r}\in \Omega _{pi}$ and $\vec{r} ^{\prime }\in
\Omega _{qj}$.

In the case of visible light the wave vector dependence of the
permittivity is negligeable. \cite{RS90} Therefore, after having
substituted Eqs.\ (\ref{eq:lhsmateq}) and (\ref{eq:rhsmateq}) into
Eq.\ (\ref{eq:intmateqrw}), only the first term in the power series
expansions has to be kept, which immediately leads to
\[
\sum_{p=1}^{N}\left[ {\cal \vec{D}}_{p}-\sum_{q=1}^{N}\tilde{\bbox{\epsilon}}%
^{pq}\left( \omega \right) {\cal \vec{E}}_{q}\right] \sum_{i}\Omega _{pi}=0\
,
\]
where the inter-- ($p\neq q$), intra--layer ($p=q$) permittivities are
given by
\[
\tilde{\bbox{\epsilon}}^{pq}\left( \omega \right) =\frac{\left( 4\pi \right)
^{2}}{\sum_{i}\Omega _{pi}}\sum_{i,j}\int_{0}^{S_{pi}}dr\
r^{2}\int_{0}^{S_{qj}}dr^{\prime }\ \left( r^{\prime }\right) ^{2}\ \tilde{%
\bbox{\epsilon}}^{pi,qj}\left( \omega ;r,r^{\prime }\right) \ .
\]
It should be noted that a similar result connecting the static current
in layer $p$ to the electric field in layer $q$ is already known from
electric transport theory in inhomogeneous \cite{BZN+94} or layered
systems \cite {WLB+96}. By using the relation ${\cal \vec{D}}_{p}=
\bbox{\tilde{\epsilon}}^{p}\left( \omega \right) {\cal \vec{E}}_{p}$,
the layer--resolved permittivities $\tilde{\bbox{\epsilon}}^{p}(\omega
)$ are then solutions of the following system of equations:
\begin{equation}
\tilde{\bbox{\epsilon}}^{p}\left( \omega \right) {\cal \vec{E}}_{p} = 
\sum_{q=1}^{N}\tilde{\bbox{\epsilon}}^{pq}\left( \omega \right) \ 
{\cal\vec{E}}_{q}\ , \qquad
p=1,\ldots,N \ .  \label{eq:defpermlp}
\end{equation}
By mapping the inter-- and intra--layer contributions
$\tilde{\bbox{\sigma}}^{pq}\left( \omega \right)$ to the
microscopically exact optical conductivity tensor
$\tilde{\bbox{\sigma}}\left( \omega \right)$, Eq.\ 
(\ref{eq:totcmpxoptcond}), onto the corresponding contributions of the
permittivity tensor
\begin{equation}
\tilde{\bbox{\epsilon}}^{pq}\left( \omega \right) =\frac{1}{N}\left[ 1-
\frac{4\pi i}{\omega }\tilde{\bbox{\sigma}}^{pq}\left( \omega \right) \right]
\ ,  \label{eq:epspq}
\end{equation}
one then can establish a well--defined macroscopical model for the
evaluation of Kerr spectra.

%
%%%%%%%%%%%%%%%%%%%%%%%%%%%%%%%%%%%%%%%%%%%%%%%%%%%%%%%%%%%%%%%%%%%%%%%
%   PARTICULARIZED 2x2 MATRIX TECHNIQUE
%%%%%%%%%%%%%%%%%%%%%%%%%%%%%%%%%%%%%%%%%%%%%%%%%%%%%%%%%%%%%%%%%%%%%%%
%

\subsection{The $2\times 2$ matrix technique}

\label{sect:2x2}

%
%%%%%%%%%%%%%%%%%%%%%%%%%%%%%%%%%%%%%%%%%%%%%%%%%%%%%%%%%%%%%%%%%%%%%%%
%   MULTIPLE REFLECTIONS AND OPTICAL INTERFERENCES
%%%%%%%%%%%%%%%%%%%%%%%%%%%%%%%%%%%%%%%%%%%%%%%%%%%%%%%%%%%%%%%%%%%%%%%
%

\subsubsection{Multiple reflections and optical interferences}

\label{sect:multref}

In contrast to the two--media approach the inclusion of all optical
reflections and interferences within a multilayer system assumes that
each layer acts as a homogeneous, anisotropic medium between two
boundaries and is characterized by a layer--resolved dielectric tensor
$\tilde{\bbox{\epsilon}}^{p}$ ($p=1,\ldots,N$). \cite{Man90,Man95}

As a first step the Fresnel or characteristic equation:\, \cite{LL99} 
\begin{equation}
\left| n_{p}^{2}\delta _{\mu \nu }-n_{p\mu }n_{p\nu }-\tilde{\epsilon}%
_{\mu \nu }^{p}\right| =0\qquad \left( \mu ,\nu ={\rm x},{\rm y},{\rm z}%
\right) \   \label{eq:chareq}
\end{equation}
has to be solved in order to determine the normal modes of the
electromagnetic waves in a particular layer $p$. \cite{ZK97} Then by
solving the Helmholtz equation for each normal modes,\, \cite{ZK97}
\begin{equation}
\sum_{\nu }\left( n_{p}^{2}\delta _{\mu \nu }-n_{p\mu }n_{p\nu }-\tilde{%
\epsilon}_{\mu \nu }^{p}\right) {\cal E}_{p\nu }=0\qquad \left( \mu ,\nu =%
{\rm x},{\rm y},{\rm z}\right) \ ,  \label{eq:pwphelm}
\end{equation}
the corresponding ${\cal E}_{p\nu }$ components of the electric field
in layer $p$ are deduced. After having obtained the ${\cal E}_{p\nu
  }$s , the curl Maxwell equation \cite{Man90,Man95}
\begin{equation}
{\cal \vec{H}}_{p}=\vec{n}_{p}\times {\cal \vec{E}}_{p}\ ,
\label{eq:pwpcurl}
\end{equation}
provides the amplitudes of the magnetic fields ${\cal \vec{H}}_{p}$
for each normal mode in layer $p$. Here the Gaussian system of units
has been used, $\vec{n}_{p}$ is the refraction vector, as given by
Eq.\ (\ref {eq:defrefvec}), $\mid\vec{n}_{p}\mid = \tilde{n}_{p}$,
which in an anisotropic medium is direction and frequency
dependent. \cite{LL99}

Finally, continuity of the tangential components of the electric and
the magnetic field at the boundary between adjacent layers leads to a
set of equations, which has to be solved recursively in order to
determine the magneto--optical coefficients of the layered system,
such as, e.g., the surface reflectivity. If no symmetry reduced
quantities $\tilde{\bbox{\epsilon}}^{p}$ are used all the previous
steps, Eqs.\ (\ref{eq:chareq})--(\ref{eq:pwpcurl}), have to be
performed numerically by using -- for example -- the $2\times 2$
matrix technique of Mansuripur. \cite{Man90,Man95}

Most frequently MOKE experiments are performed in polar geometry using
normal incidence. Therefore in the following the $2\times 2$ matrix
technique of Mansuripur is confined to this particular experimental
geometry. This reduction has the advantage that with the exception of
the last step in which the surface reflectivity has to be evaluated,
all the other steps can be carried out analytically.

In the case of cubic, hexagonal or tetragonal systems and the
orientation of the magnetization $\vec{M}_{p}$ pointing along the
surface normal (${\rm z}$--direction), the layer--resolved
permittivity tensor is given by
\begin{equation}
\tilde{\bbox{\epsilon}}^{p}=\left( 
\begin{array}{lll}
\ \ \tilde{\epsilon}_{{\rm xx}}^{p} & \ \ \tilde{\epsilon}_{{\rm xy}}^{p} & 
\ \ 0 \\ 
-\tilde{\epsilon}_{{\rm xy}}^{p} & \ \ \tilde{\epsilon}_{{\rm xx}}^{p} & \ \
0 \\ 
\ \ 0 & \ \ 0 & \ \ \tilde{\epsilon}_{{\rm zz}}^{p}
\end{array}
\right) \ .  \label{eq:eps4mparzapp}
\end{equation}
Assuming that $\tilde{\epsilon}_{{\rm zz}}^{p}\simeq
\tilde{\epsilon}_{{\rm xx}}^{p}$ ($p=1,\ldots ,N$), the error
introduced by this simplification -- as easily can be shown -- is
proportional to the difference $\tilde{\epsilon}_{{\rm
zz}}^{p}-\tilde{\epsilon}_{{\rm xx}}^{p}$, which in turn is usually
small enough to be neglected. If in polar geometry the incidence is
normal,\
\[
\tilde{n}_{p{\rm x}}=\tilde{n}_{p{\rm y}}=0\quad \mbox{for}\quad p=1,\ldots
,N\ , 
\]
the characteristic equation (\ref{eq:chareq}) provides four normal
modes of electromagnetic waves in a layer $p$:
\[
\tilde{n}_{p{\rm z}}=\pm \sqrt{\tilde{\epsilon}_{{\rm xx}}^{p}\pm i\tilde{%
\epsilon}_{{\rm xy}}^{p}}\ . 
\]
Two of these four solutions are always situated in the lower half and
the other two in the upper half of the complex plane. The first two
solutions, $\tilde{n}_{p{\rm z}}^{(1)}$ and $\tilde{n}_{p{\rm z}
  }^{(2)}$, correspond to a ``downward'' (negative ${\rm z}$
--direction) and the other two, $\tilde{n}_{p{\rm z}}^{(3)}$ and
$\tilde{n} _{p{\rm z}}^{(4)}$, an ``upward'' (positive ${\rm
  z}$--direction) propagation of the electromagnetic waves.
\cite{Man90,Man95} These two different kinds of cases are given by
\begin{equation}
\left\{ 
\begin{array}{lll}
\tilde{n}_{p{\rm z}}^{(1)} & = & -\sqrt{\tilde{\epsilon}_{{\rm xx}}^{p}+i%
\tilde{\epsilon}_{{\rm xy}}^{p}} \\ 
\tilde{n}_{p{\rm z}}^{(2)} & = & -\sqrt{\tilde{\epsilon}_{{\rm xx}}^{p}-i%
\tilde{\epsilon}_{{\rm xy}}^{p}}
\end{array}
\right.  \label{eq:soldw4polgeomappni}
\end{equation}
and 
\begin{equation}
\left\{ 
\begin{array}{lll}
\tilde{n}_{p{\rm z}}^{(3)} & = & \sqrt{\tilde{\epsilon}_{{\rm xx}}^{p}+i%
\tilde{\epsilon}_{{\rm xy}}^{p}} \\ 
\tilde{n}_{p{\rm z}}^{(4)} & = & \sqrt{\tilde{\epsilon}_{{\rm xx}}^{p}-i%
\tilde{\epsilon}_{{\rm xy}}^{p}}
\end{array}
\right. \ . \label{eq:solup4polgeomappni}
\end{equation}

If in a given multilayer system a particular layer $p$ is
paramagnetic, its permittivity tensor $\tilde{\bbox{\epsilon}}^{p}$ is
again of form shown in Eq.\ (\ref{eq:eps4mparzapp}), with
$\tilde{\epsilon}_{{\rm xy}}^{p}=0$ and $\tilde{\epsilon}_{{\rm
    zz}}^{p} = \tilde{\epsilon}_{{\rm xx}}^{p}$.  In this case only
two beams are propagating, namely those characterized by $
\tilde{n}_{p{\rm z}}^{(1)}\equiv \tilde{n}_{p{\rm
    z}}^{(2)}=-\sqrt{\tilde{ \epsilon}_{{\rm xx}}^{p}}$ and
$\tilde{n}_{p{\rm z}}^{(3)}\equiv \tilde{n} _{p{\rm
    z}}^{(4)}=\sqrt{\tilde{\epsilon}_{{\rm xx}}^{p}}$.  Furthermore,
since the vacuum is a homogeneous, isotropic, semi--infinite medium,
in addition to $\tilde{\epsilon}_{{\rm xy}}=0$,
$\tilde{\epsilon}_{{\rm xx}}=\tilde{\epsilon}_{{\rm zz}}=1$.

For each solution $n_{p{\rm z}}^{(k)}$ ($k=1,\ldots ,4$)\ of the
characteristic equation (\ref{eq:chareq}), the electric field must
satisfy the Helmholtz equation (\ref{eq:pwphelm}). Because not all of
the equations are independent, these can be solved only for two
components of the electric field keeping the third one arbitrary.
Therefore, following the strategy proposed by Mansuripur, for beam 1
($n_{p {\rm z}}^{(1)}$) and beam 3 ($n_{p{\rm z}}^{(3)}$), the
corresponding ${\cal E}_{p{\rm x}}^{(k)}$ are chosen arbitrary,
whereas for beam 2 ($n_{p{\rm z}}^{(2)}$) and beam 4 ($n_{p{\rm
    z}}^{(4)}$), the ${\cal E}_{p{\rm y}}^{(k)}$ are arbitrary.
\cite{Man90,Man95} For polar geometry and normal incidence, the
solutions of the Helmholtz equation (\ref{eq:pwphelm}) are given in
Table\ \ref{tab:solhelm4polgeomappni} and the corresponding components
of the magnetic field as obtained from Eq.\ (\ref{eq:pwpcurl}) are
listed in Table\ \ref{tab:solcurl4polgeomappni}.

%
%%%%%%%%%%%%%%%%%%%%%%%%%%%%%%%%%%%%%%%%%%%%%%%%%%%%%%%%%%%%%%%%%%%%%%%
%   Table: SOLUTIONS OF THE HELMHOLTZ EQUATION
%%%%%%%%%%%%%%%%%%%%%%%%%%%%%%%%%%%%%%%%%%%%%%%%%%%%%%%%%%%%%%%%%%%%%%%
%
%
%%%%%%%%%%%%%%%%%%%%%%%%%%%%%%%%%%%%%%%%%%%%%%%%%%%%%%%%%%%%%%%%%%%%%%%
%   Table: SOLUTIONS OF THE CURL MAXWELL EQUATION
%%%%%%%%%%%%%%%%%%%%%%%%%%%%%%%%%%%%%%%%%%%%%%%%%%%%%%%%%%%%%%%%%%%%%%%
%

%
%%%%%%%%%%%%%%%%%%%%%%%%%%%%%%%%%%%%%%%%%%%%%%%%%%%%%%%%%%%%%%%%%%%%%%%
%   LAYER--RESOLVED REFLECTIVITY MATRIX
%%%%%%%%%%%%%%%%%%%%%%%%%%%%%%%%%%%%%%%%%%%%%%%%%%%%%%%%%%%%%%%%%%%%%%%
%

\subsubsection{Layer--resolved reflectivity matrix}
\label{sect:layrefmtx}

\begin{figure}[hbtp] \centering
\includegraphics[width=0.47\columnwidth,clip]{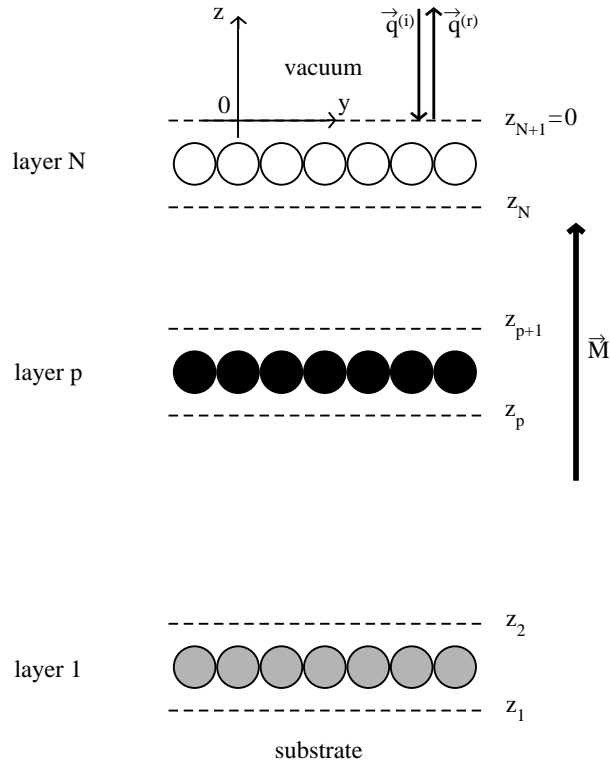}
\caption[Macroscopic model of a layer system.]
    {\label{fig:model}
      The macroscopic model used for a layered system within the
      $2\times 2$ matrix technique for polar geometry and normal
      incidence.  The ${\rm x}$ axis is perpendicular to the plane of
      the figure, $\vec{q}^{\,(i)}$ is the incident and
      $\vec{q}^{\,(r)}$ is the reflected wave vector. $\vec{M}$
      denotes the total spontaneous magnetization of the system.}
\end{figure}

Numbering the layers starting from the first one on top of the
substrate towards the surface, the surface layer has the layer index
$p=N$, see Fig.\ \ref{fig:model}.  The $2\times 2$ reflectivity matrix
${\cal R}_{p}$ at the lower boundary $z_{p}$ of layer $p$ is given by:
\cite{Man90,Man95}
\begin{equation}
\left( 
\begin{array}{l}
{\cal E}_{p{\rm x}}^{(3)} \\ 
{\cal E}_{p{\rm y}}^{(4)}
\end{array}
\right) ={\cal R}_{p}\left( 
\begin{array}{l}
{\cal E}_{p{\rm x}}^{(1)} \\ 
{\cal E}_{p{\rm y}}^{(2)}
\end{array}
\right) =\left( 
\begin{array}{ll}
\tilde{r}_{p} & 0 \\ 
0 & \tilde{r}_{p}^{\prime }
\end{array}
\right) \left( 
\begin{array}{l}
{\cal E}_{p{\rm x}}^{(1)} \\ 
{\cal E}_{p{\rm y}}^{(2)}
\end{array}
\right) \ ,  \label{eq:defrefmtxlj}
\end{equation}
see also the explicit discussion in the Appendix. The tangential
components of the electric and magnetic field at a point $ z_{p}^{+}$
just above the boundary $z_{p}$ are then given by
\begin{equation}
\left\{ 
\begin{array}{llr}
\left( 
\begin{array}{l}
{\cal E}_{p{\rm x}} \\ 
{\cal E}_{p{\rm y}}
\end{array}
\right) _{z_{p}^{+}} & = & {\cal A}\left( {\cal I}+{\cal R}_{p}\right) 
\left( 
\begin{array}{l}
{\cal E}_{p{\rm x}}^{(1)} \\ 
{\cal E}_{p{\rm y}}^{(2)}
\end{array}
\right) \\[2ex] 
\left( 
\begin{array}{l}
{\cal H}_{p{\rm x}} \\ 
{\cal H}_{p{\rm y}}
\end{array}
\right) _{z_{p}^{+}} & = & {\cal B}_{p}^{12}\left( {\cal I} -{\cal R}_{p}\right) 
\left( 
\begin{array}{l}
{\cal E}_{p{\rm x}}^{(1)} \\ 
{\cal E}_{p{\rm y}}^{(2)}
\end{array}
\right)
\end{array}
\right. \ ,  \label{eq:tanEHlwbound}
\end{equation}
where according to Tables\ \ref{tab:solhelm4polgeomappni} and \ref
{tab:solcurl4polgeomappni}, 
\begin{equation}
{\cal A}\equiv \left( 
\begin{array}{cc}
1 & i \\ 
i & 1
\end{array}
\right) \ ,\qquad {\cal B}_{p}^{12}=\left( 
\begin{array}{rr}
-i\tilde{n}_{p{\rm z}}^{(1)} & -\tilde{n}_{p{\rm z}}^{(2)} \\ 
\tilde{n}_{p{\rm z}}^{(1)} & i\tilde{n}_{p{\rm z}}^{(2)}
\end{array}
\right)  \label{eq:ABmtx}
\end{equation}
and ${\cal I}$ is the $2\times 2$ unit matrix.

Using the lower boundary $z_{p-1}$ as reference plane for the four
beams in layer $p-1$, the tangential components of the electric and
magnetic field at a point $z_{p}^{-}$ just below the boundary $z_{p}$
are of the form
\begin{equation}
\left\{ 
\begin{array}{llr}
\left( 
\begin{array}{l}
{\cal E}_{p{\rm x}} \\ 
{\cal E}_{p{\rm y}}
\end{array}
\right) _{z_{p}^{-}} & = & {\cal A}\left( {\cal C}_{p-1}^{12}+{\cal C}%
_{p-1}^{34}{\cal R}_{p-1}\right) \left( 
\begin{array}{l}
{\cal E}_{p-1{\rm x}}^{(1)} \\ 
{\cal E}_{p-1{\rm y}}^{(2)}
\end{array}
\right) \\[2ex] 
\left( 
\begin{array}{l}
{\cal H}_{p{\rm x}} \\ 
{\cal H}_{p{\rm y}}
\end{array}
\right) _{z_{p}^{-}} & = & {\cal B}_{p-1}^{12}\left( {\cal C}_{p-1}^{12}-%
{\cal C}_{p-1}^{34}{\cal R}_{p-1}\right) \left( 
\begin{array}{l}
{\cal E}_{p-1{\rm x}}^{(1)} \\ 
{\cal E}_{p-1{\rm y}}^{(2)}
\end{array}
\right)
\end{array}
\right. \ ,  \label{eq:tanEHupbound}
\end{equation}
where 
\begin{equation}
{\cal C}_{p-1}^{k,k+1}\equiv \left( 
\begin{array}{cc}
e^{-i\tilde{\varphi}_{p-1}^{(k)}} & 0 \\ 
0 & e^{-i\tilde{\varphi}_{p-1}^{(k+1)}}
\end{array}
\right) \qquad k=1,3\ ,  \label{eq:Cmtx}
\end{equation}
with 
\[
\tilde{\varphi}_{p-1}^{(k)}\equiv q_{0}\tilde{n}_{p-1{\rm z}}^{(k)}d_{p-1}\
,\qquad k=1,\ldots ,4 \ .
\]
In here $d_{p}\equiv z_{p+1}-z_{p}$ is the thickness of layer $p$,
$\tilde{n}_{p-1{\rm z}}^{(k)}$ is defined in Eqs.\ 
(\ref{eq:soldw4polgeomappni} ) and (\ref{eq:solup4polgeomappni}) and
$q_{0}$ is the propagation constant in vacuum, see Sect.\ 
\ref{sect:lreps}.

Based on Eqs.\ (\ref{eq:tanEHlwbound}) and (\ref{eq:tanEHupbound}),
the continuity of the tangential components of the electric and
magnetic field on the boundary $z_{p}$ implies that
\[
\left\{ 
\begin{array}{rrr}
\left( {\cal I}+{\cal R}_{p}\right) \left( 
\begin{array}{l}
{\cal E}_{p{\rm x}}^{(1)} \\ 
{\cal E}_{p{\rm y}}^{(2)}
\end{array}
\right) & = & \left( {\cal C}_{p-1}^{12}+{\cal C}_{p-1}^{34}{\cal R}%
_{p-1}\right) \left( 
\begin{array}{l}
{\cal E}_{p-1{\rm x}}^{(1)} \\ 
{\cal E}_{p-1{\rm y}}^{(2)}
\end{array}
\right) \\[4ex] 
{\cal B}_{p}^{12}\left( {\cal I}-{\cal R}_{p}\right) \left( 
\begin{array}{l}
{\cal E}_{p{\rm x}}^{(1)} \\ 
{\cal E}_{p{\rm y}}^{(2)}
\end{array}
\right) & = & {\cal B}_{p-1}^{12}\left( {\cal C}_{p-1}^{12}-{\cal C}%
_{p-1}^{34}{\cal R}_{p-1}\right) \left( 
\begin{array}{l}
{\cal E}_{p-1{\rm x}}^{(1)} \\ 
{\cal E}_{p-1{\rm y}}^{(2)}
\end{array}
\right)
\end{array}
\right. \ , 
\]
such that by eliminating the electric field vectors, one immediately
gets
\[
{\cal D}_{p-1}\left( 1+{\cal R}_{p}\right) ={\cal B}%
_{p}^{12}\left( {\cal I}-{\cal R}_{p}\right) \ , 
\]
where 
\begin{equation}
{\cal D}_{p-1}\equiv {\cal B}_{p-1}^{12}\left( {\cal C}_{p-1}^{12}-{\cal C}%
_{p-1}^{34}{\cal R}_{p-1}\right) \left( {\cal C}_{p-1}^{12}+{\cal C}%
_{p-1}^{34}{\cal R}_{p-1}\right) ^{-1}\ . \label{eq:Dmtx}
\end{equation}
${\cal R}_{p}$ is therefore given in terms of ${\cal R}_{p-1}$ by the
following simple recursion relation:
\begin{equation}
{\cal R}_{p}=\left( {\cal B}_{p}^{12}+{\cal D}_{p-1}\right) ^{-1}\left( 
{\cal B}_{p}^{12}-{\cal D}_{p-1}\right) \qquad
p=1,\ldots,N\ .  \label{eq:Rprecur}
\end{equation}
In order to determine the reflectivity matrix ${\cal R}_{N}$ of the
surface layer, one has to evaluate all reflectivity matrices ${\cal R}
_{p}$ for all layers below the surface layer. This requires to start
the iterative procedure at the first layer ($p=1$) on top of the
substrate. But in order to calculate ${\cal R}_{1}$, one needs to know
the $2\times 2$ matrix ${\cal D}_{0}$ corresponding to the substrate,
see Eq.\ (\ref{eq:Rprecur}). This in turn, according to Eq.\ 
(\ref{eq:Dmtx}) is only the case, if the reflectivity matrix ${\cal
  R}_{0}$ of the substrate is available. In order to achieve this, one
has to formulate the tangential components of the electric and
magnetic field at $z_{1}^{-}$ by taking into account that the
substrate is a semi--infinite bulk without any boundaries and hence
${\cal R}_{0}=0$. \cite {Man90,Man95} Thus ${\cal D }_{0}={\cal
  B}_{0}^{12}$, which according to Eq.\ (\ref{eq:ABmtx}) requires to
specify the permittivity of the substrate.

%
%%%%%%%%%%%%%%%%%%%%%%%%%%%%%%%%%%%%%%%%%%%%%%%%%%%%%%%%%%%%%%%%%%%%%%%
%   SURFACE REFLECTIVITY MATRIX
%%%%%%%%%%%%%%%%%%%%%%%%%%%%%%%%%%%%%%%%%%%%%%%%%%%%%%%%%%%%%%%%%%%%%%%
%

\subsubsection{Surface reflectivity matrix}

\label{sect:surfref}

In the vacuum region, since $\tilde{\epsilon}_{{\rm xx}}=1$, $\tilde{
  \epsilon}_{{\rm xy}}=0$, one has to deal with the superposition of
only two beams, namely that of the incident and reflected
electromagnetic waves.  These beams are related through the surface
reflectivity matrix $R_{{\rm surf }}$ such that for polar geometry and
normal incidence,
\begin{equation}
\left( 
\begin{array}{l}
{\cal E}_{{\rm vac,x}}^{({\rm r})} \\ 
{\cal E}_{{\rm vac,y}}^{({\rm r})}
\end{array}
\right) = R_{{\rm surf}} \left( 
\begin{array}{l}
{\cal E}_{{\rm vac,x}}^{({\rm i})} \\ 
{\cal E}_{{\rm vac,y}}^{({\rm i})}
\end{array}
\right) \equiv \left( 
\begin{array}{rl}
\tilde{r}_{{\rm xx}} & \tilde{r}_{{\rm xy}} \\ 
-\tilde{r}_{{\rm xy}} & \tilde{r}_{{\rm xx}}
\end{array}
\right) \left( 
\begin{array}{l}
{\cal E}_{{\rm vac,x}}^{({\rm i})} \\ 
{\cal E}_{{\rm vac,y}}^{({\rm i})}
\end{array}
\right) \ ,  \label{eq:defsurfrefmtx}
\end{equation}
see the Appendix.  Thus the tangential components of the electric and
magnetic field at a point $z_{N+1}^{+}$, namely just above the
interface between the vacuum and the surface, are given by
\begin{equation}
\left\{ 
\begin{array}{llr}
\left( 
\begin{array}{l}
{\cal E}_{{\rm vac,x}} \\ 
{\cal E}_{{\rm vac,y}}
\end{array}
\right) _{z_{N+1}^{+}} & = & \left( {\cal I}+R_{{\rm surf}%
}\right) \left( 
\begin{array}{l}
{\cal E}_{{\rm vac,x}}^{({\rm i})} \\ 
{\cal E}_{{\rm vac,y}}^{({\rm i})}
\end{array}
\right) \\[2ex] 
\left( 
\begin{array}{l}
{\cal H}_{{\rm vac,x}} \\ 
{\cal H}_{{\rm vac,y}}
\end{array}
\right) _{z_{N+1}^{+}} & = & \left( B_{{\rm vac}}^{12}+B_{{\rm vac}}^{34}R_{%
{\rm surf}}\right) \left( 
\begin{array}{l}
{\cal E}_{{\rm vac,x}}^{({\rm i})} \\ 
{\cal E}_{{\rm vac,y}}^{({\rm i})}
\end{array}
\right)
\end{array}
\right. \ ,  \label{eq:tanEHvaclwbound}
\end{equation}
where 
\begin{equation}
B_{{\rm vac}}^{12}=\left( 
\begin{array}{rc}
0 & 1 \\ 
-1 & 0
\end{array}
\right) \quad \mbox{and}\quad B_{{\rm vac}}^{34}=\left( 
\begin{array}{cr}
0 & -1 \\ 
1 & 0
\end{array}
\right) \ .  \label{eq:Bmtx4vacni}
\end{equation}
According to Eqs.\ (\ref{eq:tanEHupbound}) and
(\ref{eq:tanEHvaclwbound}), the continuity of the tangential
components of the electric and magnetic fields at the vacuum and
surface layer interface, $z_{N+1}=0$, can be written as
\[
\left\{ 
\begin{array}{rlr}
\left( {\cal I}+R_{{\rm surf}}\right) \left( 
\begin{array}{l}
{\cal E}_{{\rm vac,x}}^{({\rm i})} \\ 
{\cal E}_{{\rm vac,y}}^{({\rm i})}
\end{array}
\right) & = & {\cal A}\left( {\cal C}_{N}^{12}+{\cal C}_{N}^{34}{\cal R}%
_{N}\right) \left( 
\begin{array}{l}
{\cal E}_{N{\rm x}}^{(1)} \\ 
{\cal E}_{N{\rm y}}^{(2)}
\end{array}
\right) \\[4ex] 
\left( B_{{\rm vac}}^{12}+B_{{\rm vac}}^{34}R_{{\rm surf}}\right) \left( 
\begin{array}{l}
{\cal E}_{{\rm vac,x}}^{({\rm i})} \\ 
{\cal E}_{{\rm vac,y}}^{({\rm i})}
\end{array}
\right) & = & {\cal B}_{N}^{12}\left( {\cal C}_{N}^{12}-{\cal C}_{N}^{34}%
{\cal R}_{N}\right) \left( 
\begin{array}{l}
{\cal E}_{N{\rm x}}^{(1)} \\ 
{\cal E}_{N{\rm y}}^{(2)}
\end{array}
\right)
\end{array}
\right. \ . 
\]
By eliminating from this system of equations the electric field
vectors, it follows that
\[
{\cal F}_{N}\left( 1+R_{{\rm surf}}\right) =B_{{\rm vac}%
}^{12}+B_{{\rm vac}}^{34}R_{{\rm surf}}\ , 
\]
where 
\begin{equation}
{\cal F}_{N}\equiv {\cal B}_{N}^{12}\left( {\cal C}_{N}^{12}-{\cal C}%
_{N}^{34}{\cal R}_{N}\right) \left( {\cal C}_{N}^{12}+{\cal C}_{N}^{34}{\cal %
R}_{N}\right) ^{-1}{\cal A}^{-1}={\cal D}_{N}{\cal A}^{-1}\ .
\label{eq:Fmtx}
\end{equation}
Thus for the surface reflectivity matrix one obtains
\begin{equation}
R_{{\rm surf}}=\left( {\cal F}_{N}-B_{{\rm vac}}^{34}\right) ^{-1}\left( B_{%
{\rm vac}}^{12}-{\cal F}_{N}\right) \ .  \label{eq:Rsurf}
\end{equation}
The surface reflectivity matrix $R_{{\rm surf}}$ is therefore of the
form given in Eq.\ (\ref{eq:defsurfrefmtx}), see also Appendix.  In
spherical coordinates, one immediately obtains the complex
reflectivity of the right-- and left--handed circularly polarized
light as
\[
\tilde{r}_{\pm }=\tilde{r}_{{\rm xx}}\mp i\tilde{r}_{{\rm xy}}\ , 
\]
which in turn determines the Kerr rotation angle $\theta _{{\rm K}}$
and ellipticity $\epsilon _{{\rm K}}$, see Eqs.\ (\ref{eq:exactkang})
and (\ref {eq:exactkell}).

%
%%%%%%%%%%%%%%%%%%%%%%%%%%%%%%%%%%%%%%%%%%%%%%%%%%%%%%%%%%%%%%%%%%%%%%%
%   SELF-CONSISTENT LAYER-RESOLVED PERMITTIVITY
%%%%%%%%%%%%%%%%%%%%%%%%%%%%%%%%%%%%%%%%%%%%%%%%%%%%%%%%%%%%%%%%%%%%%%%
%

\subsubsection{Self--consistent layer--resolved permittivities}

\label{sect:sclreps}

In order to calculate for a homogeneous, anisotropic layer $p$ the
corresponding dielectric tensor (\ref{eq:eps4mparzapp}) from the
inter-- and intra--layer permittivities defined in Eq.\
(\ref{eq:epspq}), the following linear system of equations
\[
\left( 
\begin{array}{rr}
\tilde{\epsilon}_{{\rm xx}}^{p} & \tilde{\epsilon}_{{\rm xy}}^{p} \\ 
-\tilde{\epsilon}_{{\rm xy}}^{p} & \tilde{\epsilon}_{{\rm xx}}^{p}
\end{array}
\right) \left( 
\begin{array}{l}
{\cal E}_{p{\rm x}} \\ 
{\cal E}_{p{\rm y}}
\end{array}
\right) =\sum_{q=1}^{N}\left( 
\begin{array}{rr}
\tilde{\epsilon}_{{\rm xx}}^{pq} \  & \tilde{\epsilon}_{{\rm xy}}^{pq}
\  \\ 
-\tilde{\epsilon}_{{\rm xy}}^{pq} \  & \tilde{\epsilon}_{{\rm xx}%
}^{pq} \ 
\end{array}
\right) \left( 
\begin{array}{l}
{\cal E}_{q{\rm x}} \\ 
{\cal E}_{q{\rm y}}
\end{array}
\right) 
\]
has to be solved, see Eq.\ (\ref{eq:defpermlp}).  Here for ${\cal
\vec{E}}_{p}$ one can take the following Ansatz:
\[
\left( 
\begin{array}{l}
{\cal E}_{p{\rm x}} \\ 
{\cal E}_{p{\rm y}}
\end{array}
\right) \equiv \left( 
\begin{array}{l}
{\cal E}_{p{\rm x}} \\ 
{\cal E}_{p{\rm y}}
\end{array}
\right) _{z=z_{p}^{+}+\frac{d_{p}}{2}}={\cal A}\left[ \left( {\cal C}%
_{p}^{12}\right) ^{\frac{1}{2}}+\left( {\cal C}_{p}^{34}\right) ^{\frac{1}{2}%
}{\cal R}_{p}\right] \left( 
\begin{array}{l}
{\cal E}_{p{\rm x}}^{(1)} \\ 
{\cal E}_{p{\rm y}}^{(2)}
\end{array}
\right) \ , 
\]
where due to Eq.\ (\ref{eq:Cmtx}), 
\[
\left( {\cal C}_{p}^{k,k+1}\right) ^{\frac{1}{2}}\equiv \left( 
\begin{array}{cc}
e^{-iq_{0}\tilde{n}_{p{\rm z}}^{(k)}\frac{d_{p}}{2}} & 0 \\ 
0 & e^{-iq_{0}\tilde{n}_{p{\rm z}}^{\left( k+1\right) }\frac{d_{p}}{2}}
\end{array}
\right) \qquad k=1,3\ . 
\]
By using the continuity equation of the tangential components of the
electric field at the boundaries, see Eqs. (\ref{eq:tanEHlwbound}),
(\ref {eq:tanEHupbound}) and (\ref{eq:tanEHvaclwbound}), one then
obtains the layer--resolved permittivities as a weighted sum of the
inter-- and intra--layer permittivities defined in Eq.\
(\ref{eq:epspq}):
\begin{equation}
\left( 
\begin{array}{rr}
\tilde{\epsilon}_{{\rm xx}}^{\,p} & \tilde{\epsilon}_{{\rm xy}}^{\,p} \\%
[1ex] 
-\tilde{\epsilon}_{{\rm xy}}^{\,p} & \tilde{\epsilon}_{{\rm xx}}^{\,p}
\end{array}
\right) =\sum_{q=1}^{N}W_{pq} \, \tilde{\bbox{\epsilon}}^{\,pq} \ .
\label{eq:epspsc4mparzapp}
\end{equation}
\ \\[0.5ex]
where  
\begin{equation}
W_{pq} ={\cal A}\left( \prod_{k=0}^{N-q}{\cal W}_{q+k}\right) \left(
\prod_{k=0}^{N-p}{\cal W}_{p+k}\right) ^{-1}{\cal A}^{-1}\ ,  \label{eq:Wmtx}
\end{equation}
with
\[
{\cal W}_{p+k}=\left( {\cal I}+{\cal R}_{p+k}\right) \left( {\cal %
C}_{p+k}^{12}+{\cal C}_{p+k}^{34}{\cal R}_{p+k}\right) ^{-1}
\ , \qquad k=1,\ldots ,N-p\ , 
\]
and 
\[
{\cal W}_{p}=\left[ \left( {\cal C}_{p}^{12}\right) ^{\frac{1}{2}}+\left( 
{\cal C}_{p}^{34}\right) ^{\frac{1}{2}}{\cal R}_{p}\right] \left( {\cal C}%
_{p}^{12}+{\cal C}_{p}^{34}{\cal R}_{p}\right) ^{-1}\ , \qquad k=0. 
\]
Because the $2\times 2$ matrices ${\cal W}_{p+k}$ contain ${\cal
R}_{p}$, ${\cal C}_{p}^{12}$ and ${\cal C}_{p}^{34}$, which in turn
depend on layer--resolved permittivities $\tilde{\epsilon }_{\mu \nu
}^{p}\left( \omega \right) $, Eq.\ (\ref{eq:epspsc4mparzapp}) has to
be solved iteratively.

The self--consistent procedure can be started by putting all $2\times
2$ weighting matrices $W_{pq} $ in Eq.$\ $(\ref{eq:Wmtx}) to unity,
i.e., by neglecting the phase differences of the electromagnetic waves
between the lower and upper boundaries in each layer $p$,
\begin{equation}
\tilde{\epsilon}_{\mu \nu }^{\,p}\left( \omega \right) ^{(0)}=\sum_{q=1}^{N}%
\tilde{\epsilon}_{\mu \nu }^{\,pq}\left( \omega \right) \ .
\label{eq:eps1stguess}
\end{equation}
These quantities $\tilde{\epsilon}_{\mu \nu }^{\,p}\left( \omega
\right) ^{(0)}$ can be used to calculate ${\cal R}_{p}^{(0)}$ in terms
of Eqs.\ (\ref{eq:Dmtx}) and (\ref{eq:Rprecur}). Improved
layer--resolved permittivities follow then from Eq.\ 
(\ref{eq:epspsc4mparzapp}). This iterative procedure has to be
repeated until the difference in the old and new layer--resolved
permittivity of layer $p$ is below a numerical threshold $\epsilon
_{p}$,
\begin{equation}
\max \ \left| \,\tilde{\epsilon}_{\mu \nu }^{\,p}\left( \omega \right)
^{(i+1)}-\tilde{\epsilon}_{\mu \nu }^{\,p}\left( \omega \right)
^{(i)}\,\right| \leq \varepsilon _{p}\ . 
\label{eq:epspconv}
\end{equation}

%
%%%%%%%%%%%%%%%%%%%%%%%%%%%%%%%%%%%%%%%%%%%%%%%%%%%%%%%%%%%%%%%%%%%%%%%
%   RESULTS AND DISCUSSIONS
%%%%%%%%%%%%%%%%%%%%%%%%%%%%%%%%%%%%%%%%%%%%%%%%%%%%%%%%%%%%%%%%%%%%%%%
%

\section{Results and discussions}

\label{sect:res}

>From experiments is known that Pt substrates prefer an fcc(111)
orientation. \cite{WBG+92} Therefore in the present contribution, the
calculations for the layered systems Co$\mid$Pt$_5\!\mid$Pt(111) and
Pt$_3\mid$Co$\mid$Pt$_5\!\mid$Pt(111) have been performed, with five
Pt layers serving as buffer \cite{PZU+99} to bulk fcc Pt.

The Fermi level in Eqs.\ (\ref{eq:Sgmw}) and (\ref{eq:Sgm0}) is that
of paramagnetic fcc Pt bulk (lattice parameter of 7.4137 a.u.), which
also serves as parent lattice, \cite{PZU+99} i.e., no layer relaxation
is considered.

%
%%%%%%%%%%%%%%%%%%%%%%%%%%%%%%%%%%%%%%%%%%%%%%%%%%%%%%%%%%%%%%%%%%%%%%%
%   PARAMAGNETIC FCC-PT SUBSTRATE
%%%%%%%%%%%%%%%%%%%%%%%%%%%%%%%%%%%%%%%%%%%%%%%%%%%%%%%%%%%%%%%%%%%%%%%
%

\subsection{Paramagnetic fcc(111) Pt substrate}

\label{sect:substr}

\begin{figure}[hbtp] \centering
\includegraphics[width=0.47\columnwidth,clip]{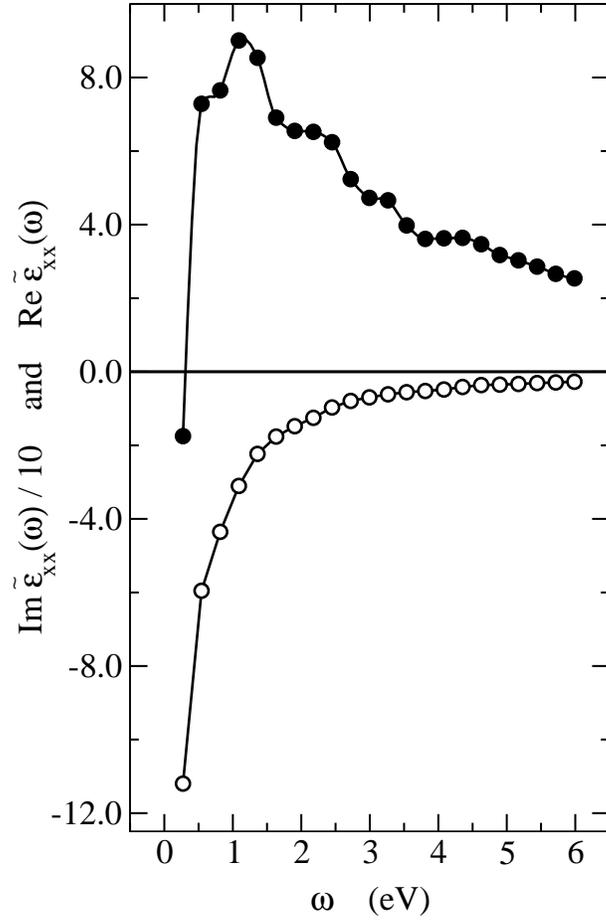}
\caption[Fcc--Pt bulk permittivities.]
    {\label{fig:Eps-ptb}
      The permittivity for fcc(111)--Pt bulk as a function of the
      photon energy $\omega$.  The real part of the permittivity is
      denoted by full circles, the imaginary part by open circles.}
\end{figure}

As mentioned above in order to determine the surface reflectivity the
permittivity tensor of the semi--infinite substrate has to be
evaluated.  As can be seen from Fig.\ \ref{fig:Eps-ptb}, the ${\rm
  xx}$--element of the permittivity tensor of the fcc(111) Pt substrate
shows a rather simple photon energy dependence. The real part of the
permittivity $\tilde{\epsilon}_{\rm xx}(\omega)$ has a peak around 1
eV, while the imaginary part of $\tilde{\epsilon}_{\rm xx}(\omega)$
exhibits an almost perfect hyperbolic frequency dependence. The strong
decay of $\tilde{\epsilon}_{\rm xx}(\omega)$ for photon energies in
the vicinity of the static limit ($\omega = 0$) can be easily
understood in terms of the Eqs.\ (\ref{eq:epspq}) and
(\ref{eq:eps1stguess}), see also Ref.\ \onlinecite{VSW01b} : for
$\omega\rightarrow 0$ the real part of $\tilde{\epsilon}_{\rm
  xx}(\omega)$ must tend to minus infinity whereas the imaginary part
has to decrease.

The ${\rm xy}$--element of the permittivity tensor for fcc--Pt(111) is
identical zero over the whole range of optical frequencies, a
functional behavior that of course does not need to be illustrated.

%
%%%%%%%%%%%%%%%%%%%%%%%%%%%%%%%%%%%%%%%%%%%%%%%%%%%%%%%%%%%%%%%%%%%%%%%
%   SELF--CONSISTENT LAYER--RESOLVED PERMITTIVITIES
%%%%%%%%%%%%%%%%%%%%%%%%%%%%%%%%%%%%%%%%%%%%%%%%%%%%%%%%%%%%%%%%%%%%%%%
%

\subsection{Self--consistent layer--resolved permittivities}

\label{sect:scperm}

In terms of the substrate and the zeroth order layer--resolved
permittivities, see Eq.\ (\ref{eq:eps1stguess}), the iterative
determination of the surface reflectivity matrix described above,
provides also self--consistent, layer--resolved permittivities
$\tilde{\epsilon}^p_{\rm xx}(\omega)$ in a very efficient manner: in
less than five iterations an accuracy of $\varepsilon_p = 10^{-13}$
for each layer $p$, see Eq.\ (\ref{eq:epspconv}), can be achieved.

\begin{figure}[hbtp] \centering
\includegraphics[width=0.47\columnwidth,clip]{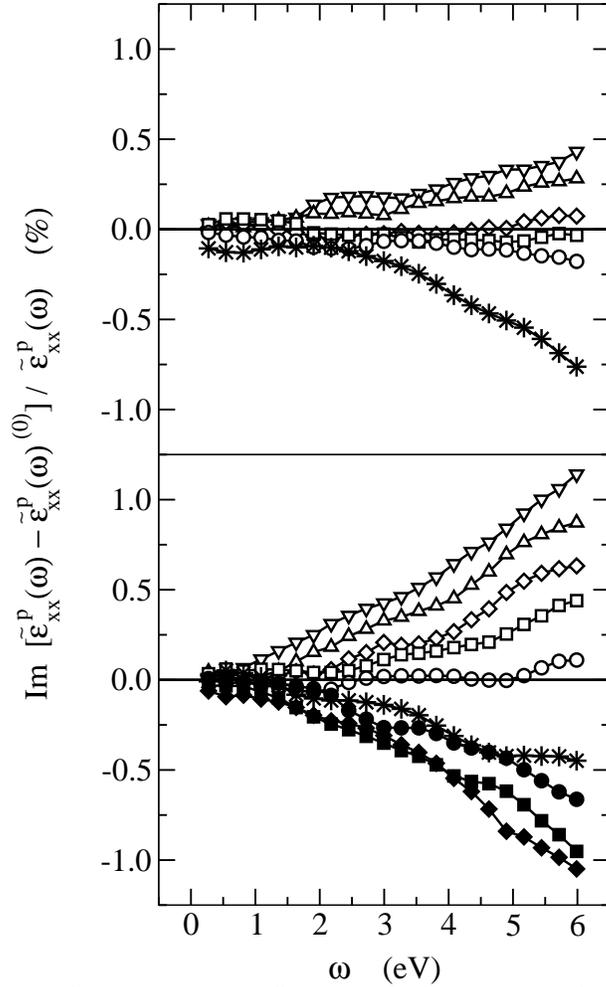} 
\caption[Relative differences in the layer--resolved permittivity.]
    {\label{fig:reldifEpsp}
      Imaginary part of the relative difference between the
      self--consistent and zeroth order layer--resolved xx--element of
      the permittivity tensor as a function of the photon energy
      $\omega$ for fcc Co$\mid$Pt$_{5}\!\mid$Pt(111) (top) and
      Pt$_{3}\!\mid$Co$\mid$Pt$_{5}\!\mid$Pt(111) (bottom).  The data
      represented by full (open) circles correspond to the first,
      squares to the second and diamonds to the third Pt layer on top
      of (under) the Co layer (stars). Open triangles down (up) denote
      the first (second) Pt layer data on top of a paramagnetic
      fcc--Pt(111) substrate.}
\end{figure}

In order to illustrate this procedure, in Fig.\ \ref{fig:reldifEpsp}
the imaginary part of the relative difference between the
self--consistent and zeroth order layer--resolved ${\rm xx}$--element
of the permittivity tensor for Co$\mid$Pt$_5\!\mid$Pt(111) with and
without Pt cap layers is displayed.

This relative difference is to be viewed as the relative error made by
using according to Eq.\ (\ref{eq:eps1stguess}) the $2\times 2$ matrix
technique with zeroth order layer--resolved permittivities.  As can be
seen from Fig.\ \ref{fig:reldifEpsp}, this relative error is layer,
frequency and system dependent. The higher the photon energy and the
bigger the layered system the less exact are the zeroth order
layer--resolved permittivities. However, for relatively small layered
systems, the relative error made by using only zeroth order
permittivities is typically below 5 \% for $\tilde{\epsilon}_{\rm
  xx}(\omega)$ and less than 20 \% for $\tilde{\epsilon}_{\rm
  xy}(\omega)$. However, the resulting relative error in the Kerr
rotation angle and the ellipticity as calculated by comparing the
spectra corresponding to the self--consistent and to the zeroth order
layer--resolved permittivities, is always less than 1 \%. Therefore,
Eq.\ (\ref{eq:eps1stguess}) can be considered as reasonably good
approximation for layer--resolved permittivities.

%
%%%%%%%%%%%%%%%%%%%%%%%%%%%%%%%%%%%%%%%%%%%%%%%%%%%%%%%%%%%%%%%%%%%%%%%
%   POLAR MOKE FOR NORMAL INCIDENCE
%%%%%%%%%%%%%%%%%%%%%%%%%%%%%%%%%%%%%%%%%%%%%%%%%%%%%%%%%%%%%%%%%%%%%%%
%

\subsection{Polar Kerr effect for normal incidence}

\label{sect:twomedvs2x2}

The systems investigated in here refer to a Co mono--layer on the top
of a fcc--Pt(111) substrate, considering also the case of three Pt cap
layers.  As already mentioned, five Pt layers serve as buffer to the
semi--infinite host in order to ensure that the induced magnetic
moments decrease monotonously to zero into the paramagnetic Pt
substrate.

\begin{figure}[hbtp] \centering
\includegraphics[width=0.47\columnwidth,clip]{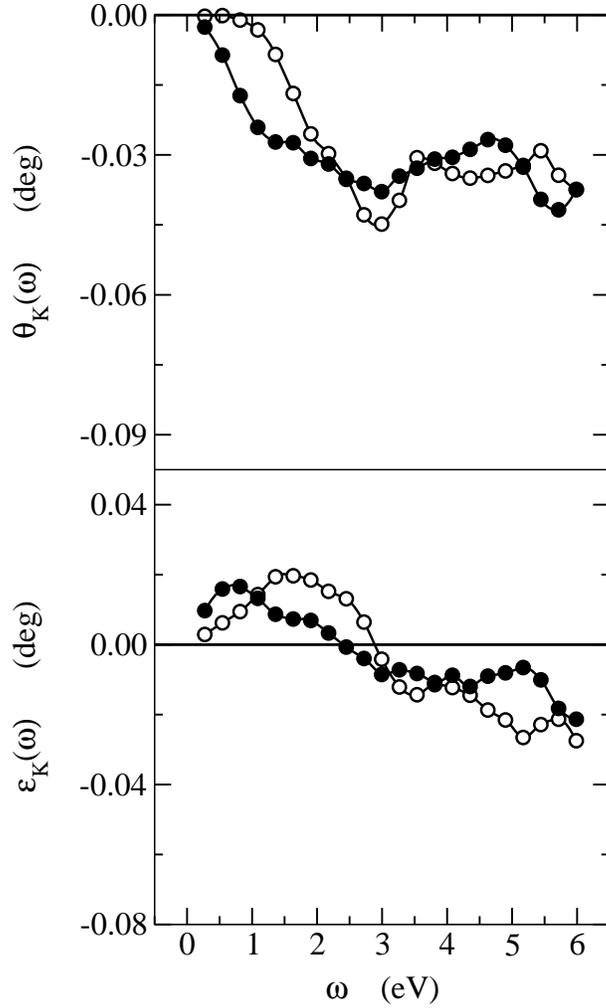} 
\caption[PMOKE 2x2 matrix technique.]
    {\label{fig:Kerr2x2} 
      The magneto--optical Kerr rotation angle
      $\theta_{\scriptstyle{\rm K}}(\omega)$ and ellipticity
      $\epsilon_{\scriptstyle{\rm K}}(\omega)$ for polar geometry and
      normal incidence as a function of the photon energy $\omega$
      obtained by applying the $2\times 2$ matrix technique for the
      self--consistent layer--resolved permittivities of fcc
      Co$\mid$Pt$_{5}\!\mid$Pt(111) (open circles) and
      Pt$_{3}\!\mid$Co$\mid$Pt$_{5}\!\mid$Pt(111) (full circles).}
\end{figure}

The ab--initio Kerr spectra obtained from self--consistent
layer--resolved permittivities by applying the $2\times 2$ matrix
technique are shown in Fig.\ \ref{fig:Kerr2x2}. Usually, in
experiments Pt cap layers are deposited on top of Co in order to
prevent the oxidation of the surface. \cite{FBT00} By performing a
separate, magnetic anisotropy calculation, \cite{PZU+99} we have found
that Co$\mid$Pt(111) exhibits perpendicular magnetization only in the
presence of Pt cap layers. Therefore, for the polar Kerr spectra of
the Co$\mid$Pt$_{5}\!\mid$Pt(111) system shown in Fig.\ 
\ref{fig:Kerr2x2} the polar geometry, namely the perpendicular
orientation of the magnetization is imposed.

\begin{figure}[hbtp] \centering
\includegraphics[width=0.47\columnwidth,clip]{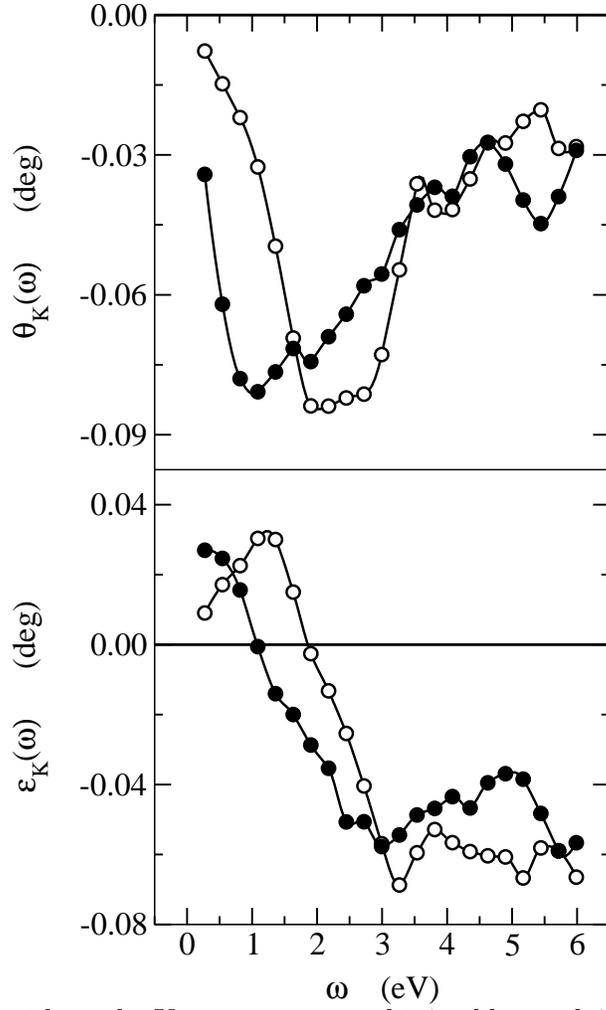} 
\caption[PMOKE two--media approach.]
    {\label{fig:Kerrstd} 
      As in Fig.\ \ref{fig:Kerr2x2}, but here the Kerr spectra was
      obtained by applying the two--media approach for fcc
      Co$\mid$Pt$_{5}\!\mid$Pt(111) (open circles) and
      Pt$_{3}\!\mid$Co$\mid$Pt$_{5}\!\mid$Pt(111) (full circles).}
\end{figure}

Analyzing in Fig.\ \ref{fig:Kerr2x2} the Kerr spectra of the capped
and that of the uncapped systems, several differences can be observed.
The negative peak in the Kerr rotation angle $\theta_{\rm K}$ at 3 eV
in the spectrum of Co$\mid$Pt$_{5}\!\mid$Pt(111) almost disappears
from the spectrum in the case of the capped layered system. In the
Kerr ellipticity the zero location at 2.8 eV observed for the uncapped
system is shifted to 2.5 eV for the capped system system and
simultaneously the infrared (IR) positive peak is shrunk and moved
towards lower photon energies. Besides these features, the sign of the
ultraviolet (UV) peak in both, the Kerr rotation and ellipticity
spectra is changed, when the Co surface layer is capped.  This
particular feature can also be observed in the Kerr spectra obtained
by using the two--media approach, see Fig.\ \ref{fig:Kerrstd}. In the
two--media Kerr rotation spectrum, the negative IR peak at 2 eV in
Co$\mid$Pt$_{5}\!\mid$Pt(111) is shifted to about 1 eV in case of the
capped system.

Comparing the spectra in Fig.\ \ref{fig:Kerr2x2} with those in Fig.\ 
\ref{fig:Kerrstd}, it is evident that the theoretical Kerr spectra
depend indeed on the macroscopic model used to describe the
propagation of electromagnetic waves in the system.  Because the
systems investigated in here are much smaller than those used in
experiments, \cite{FBT00,UUY+96} a strict quantitative comparison with
experimental data cannot be made. However, a qualitative comparison
based on the well--known, general features of the Co$\mid$Pt
experimental Kerr spectra is still possible: \cite{UUY+96} the Kerr
rotation angle shows (a) a small negative IR peak at 1.5 eV, which
decreases in amplitude with decreasing Co thickness, and (b) a high
and broad negative UV peak, which moves from 4.1 eV to 3.9 eV for
increasing Co thickness. The Kerr ellipticity is characterized by (a)
a shift of the zero location at 1.5 eV (pure Co film) to 3.7 eV with
decreasing Co thickness, (b) a positive peak around 3 eV and (c) a
shift of the minimum at 4.9 eV in pure Co film towards higher photon
energies.

In the Kerr rotation spectra of the capped system shown in Fig.\ 
\ref{fig:Kerr2x2}, there is no negative IR peak around 1.5 eV, but a
negative UV exists at 5.5 eV. The Kerr ellipticity spectra of the
capped system in Fig.\ \ref{fig:Kerr2x2}, has a zero location at 2.5
eV and positive peaks show up at 0.5, 3.5, 4 and 5 eV. These features
suggest that in the case of
Pt$_{3}\!\mid$Co$\mid$Pt$_{5}\!\mid$Pt(111) the Kerr spectra obtained
by applying the $2\times 2$ matrix technique are typical for
Co$\mid$Pt layered systems.

A similar investigation of the Kerr rotation spectra in Fig.\ 
\ref{fig:Kerrstd}, reveals that for the capped system there are two
negative IR peaks at 1, 1.5 eV and a negative UV peak around 5 eV. The
Kerr ellipticity for the capped system in Fig.\ \ref{fig:Kerrstd}
shows a zero location at 1 eV (1.5 eV in case of a pure Co film), two
positive peaks at 4 and 5 eV and a small negative peak around 3 eV.
All these features make the Kerr spectra of
Pt$_{3}\!\mid$Co$\mid$Pt$_{5}\!\mid$Pt(111) described via the
two--media approach to resemble those of a pure Co film rather than
those of a Co$\mid$Pt layered system.

Previous results obtained by applying the $2\times 2$ matrix technique
using as substrate permittivity that of the last Pt--layer below the
Co one, \cite{VSW01d} have shown similar characteristics in the Kerr
spectra. Hence, these features cannot be ascribed to the pre\-sen\-ce
of the substrate. In an other contribution, \cite{VSW01c} it was shown
that the optical conductivity of these systems is dominated by the
contributions arising from the polarized Pt--layers. Therefore, the
pure Co film--like spectra obtained for
Pt$_{3}\!\mid$Co$\mid$Pt$_{5}\!\mid$Pt(111) within the two--media
approach can be seen as an indication that a layered system cannot be
approximated by a homogeneous medium in which no optical reflections
or interferences occur.

%
%%%%%%%%%%%%%%%%%%%%%%%%%%%%%%%%%%%%%%%%%%%%%%%%%%%%%%%%%%%%%%%%%%%%%%%
%   SUMMARY
%%%%%%%%%%%%%%%%%%%%%%%%%%%%%%%%%%%%%%%%%%%%%%%%%%%%%%%%%%%%%%%%%%%%%%%
%

\section{Summary}

\label{sect:summary}

We have used the $2\times 2$ matrix technique for the most frequently
used experimental set--up, namely for polar geometry and normal
incidence. This technique allows one to account for all multiple
reflections and optical interferences in a semi--infinite layered
system. The Kerr rotation angle and ellipticity can be directly
obtained from the iteratively calculated surface reflectivity matrix,
which in turn can be used to determine layer--resolved permittivities
self--consistently. For free surface of layered systems realistic
ab--initio Kerr spectra are obtained using the inter-- and
intra--layer conductivities as given by Luttinger's formula within the
spin--polarized relativistic screened Korringa--Kohn--Rostoker method.

A comparison of the theoretical Kerr spectra of
Co$\mid$Pt$_{5}\!\mid$Pt(111) and
Pt$_{3}\mid$Co$\mid$Pt$_{5}\!\mid$Pt(111) as obtained by applying the
$2\times 2$ matrix technique and the two--media approach indicates
that the former technique provides typical results for layered
systems, whereas the later approach tends to generate spectra specific
for homogeneous films on top of a substrate.

%
%%%%%%%%%%%%%%%%%%%%%%%%%%%%%%%%%%%%%%%%%%%%%%%%%%%%%%%%%%%%%%%%%%%%%%%
%   ACKNOWLEDGMENTS
%%%%%%%%%%%%%%%%%%%%%%%%%%%%%%%%%%%%%%%%%%%%%%%%%%%%%%%%%%%%%%%%%%%%%%%
%

\section*{Acknowledgments}

This work was supported by the Austrian Ministry of Science (Contract
No. 45.451), by the Hungarian National Science Foundation (Contract
No. OTKA T030240 and T029813) and partially by the RTN network
``Computational Magnetoelectronics'' (Contract
No. HPRN--CT--2000-00143).

%
%%%%%%%%%%%%%%%%%%%%%%%%%%%%%%%%%%%%%%%%%%%%%%%%%%%%%%%%%%%%%%%%%%%%%%%
%   APPENDIX
%%%%%%%%%%%%%%%%%%%%%%%%%%%%%%%%%%%%%%%%%%%%%%%%%%%%%%%%%%%%%%%%%%%%%%%
%

\section*{Appendix: symmetry of reflectivity matrices}

Since for a semi--infinite substrate, ${\cal R}_{0}=0$, ${\cal
  D}_{0}={\cal B}_{0}^{12}$, with ${\cal B}_{0}^{12}$ as given by Eq.\ 
(\ref{eq:ABmtx}), according to Eq.\ (\ref {eq:Rprecur}), the
reflectivity matrix of the first layer on top of the substrate is
given by
\[
{\cal R}_{1}=\left( 
\begin{array}{cc}
\tilde{r}_{1} & 0 \\ 
0 & \tilde{r}_{1}^{\prime }
\end{array}
\right) \ , 
\]
where 
\[
\left\{ 
\begin{array}{lll}
\tilde{r}_{1} & = & 
{\displaystyle{\tilde{n}_{1{\rm z}}^{(1)}-\tilde{n}_{0{\rm z}}^{(1)} 
\overwithdelims.. \tilde{n}_{1{\rm z}}^{(1)}+\tilde{n}_{0{\rm z}}^{(1)}}}%
\\[2ex] 
\tilde{r}_{1}^{\prime } & = & 
{\displaystyle{\tilde{n}_{1{\rm z}}^{(2)}-\tilde{n}_{0{\rm z}}^{(2)} 
\overwithdelims.. \tilde{n}_{1{\rm z}}^{(2)}+\tilde{n}_{0{\rm z}}^{(2)}}}%
\end{array}
\right. \ . 
\]

Assuming that all $p-1$ reflectivity matrices are of this diagonal
form, namely
\[
{\cal R}_{j}=\left( 
\begin{array}{cc}
\tilde{r}_{j} & 0 \\ 
0 & \tilde{r}_{j}^{\prime }
\end{array}
\right) ,\qquad j=1,\ldots ,\left( p-1\right) \ , 
\]
by taking into account Eqs.\ (\ref{eq:ABmtx}) and (\ref{eq:Cmtx}), Eq.\ (\ref
{eq:Dmtx}) immediately yields

\[
{\cal D}_{p-1}=\left( 
\begin{array}{rr}
-i\tilde{d}_{p-1} & -\tilde{d}_{p-1}^{\prime } \\ 
\tilde{d}_{p-1} & i\tilde{d}_{p-1}^{\prime }
\end{array}
\right) \ , 
\]
where 
\[
\left\{ 
\begin{array}{lll}
\tilde{d}_{p-1} & = & \tilde{n}_{p-1{\rm z}}^{(1)}%
{\displaystyle{e^{-i\tilde{\varphi}_{p-1}^{(1)}}-
e^{-i\tilde{\varphi}_{p-1}^{(3)}}\tilde{r}_{p-1} 
\overwithdelims.. e^{-i\tilde{\varphi}_{p-1}^{(1)}}+
e^{-i\tilde{\varphi}_{p-1}^{(3)}}\tilde{r}_{p-1}}}%
\\[2ex] 
\tilde{d}_{p-1}^{\prime } & = & \tilde{n}_{p-1{\rm z}}^{(2)}%
{\displaystyle{e^{-i\tilde{\varphi}_{p-1}^{(2)}}-
e^{-i\tilde{\varphi}_{p-1}^{(4)}}\tilde{r}_{p-1}^{\prime } 
\overwithdelims.. e^{-i\tilde{\varphi}_{p-1}^{(2)}}+
e^{-i\tilde{\varphi}_{p-1}^{(4)}}\tilde{r}_{p-1}^{\prime }}}%
\end{array}
\right. \ . 
\]
The reflectivity matrix of layer $p$ as obtained from the recursion
relation (\ref{eq:Rprecur}) is found to be also diagonal
\[
{\cal R}_{p}=\left( 
\begin{array}{cc}
\tilde{r}_{p} & 0 \\ 
0 & \tilde{r}_{p}^{\prime }
\end{array}
\right) \ , 
\]
where 
\[
\left\{ 
\begin{array}{lll}
\tilde{r}_{p} & = & 
{\displaystyle{\tilde{n}_{p{\rm z}}^{(1)}-\tilde{d}_{p-1} 
\overwithdelims.. \tilde{n}_{p{\rm z}}^{(1)}+\tilde{d}_{p-1}}}%
\\[2ex] 
\tilde{r}_{p}^{\prime } & = & 
{\displaystyle{\tilde{n}_{p{\rm z}}^{(2)}-\tilde{d}_{p-1}^{\prime } 
\overwithdelims.. \tilde{n}_{p{\rm z}}^{(2)}-\tilde{d}_{p-1}^{\prime }}}%
\end{array}
\right. \ , 
\]
i.e., all the layer--resolved reflectivity matrices ${\cal R}_{j}$ (
$j=1,\ldots ,N$) are diagonal matrices as was anticipated in Eq.\
(\ref{eq:defrefmtxlj}).

In terms of the diagonal reflectivity matrix of the surface layer
${\cal R}_{N}$, and ${\cal A}$ and ${\cal B}_{N}^{12}$ as given by
Eq.\ (\ref{eq:ABmtx}), Eq.\ ( \ref{eq:Fmtx}) reduces to
\[
{\cal F}_{N}=\frac{1}{2}\allowbreak \left( 
\begin{array}{cc}
i\tilde{f}_{N} & -\tilde{f}_{N}^{\prime } \\ 
\tilde{f}_{N}^{\prime } & i\tilde{f}_{N}
\end{array}
\right) \ , 
\]
where 
\[
\left\{ 
\begin{array}{lll}
\tilde{f}_{N} & = & \tilde{d}_{N}^{\prime }-\tilde{d}_{N} \\ 
\tilde{f}_{N}^{\prime } & = & \tilde{d}_{N}^{\prime }+\tilde{d}_{N}
\end{array}
\right. \ . 
\]
By using ${\cal F}_{N}$ together with the matrices defined in Eq.\
(\ref {eq:Bmtx4vacni}) in Eq.\ (\ref{eq:Rsurf}), the resulting surface
reflectivity matrix is of the form anticipated in Eq.\ (\ref
{eq:defsurfrefmtx}), i.e.\
\[
R_{{\rm surf}}=\left( 
\begin{array}{rr}
\tilde{r}_{{\rm xx}} & \tilde{r}_{{\rm xy}} \\ 
-\tilde{r}_{{\rm xy}} & \tilde{r}_{{\rm xx}}
\end{array}
\right) \ , 
\]
where
\[
\left\{ 
\begin{array}{llr}
\tilde{r}_{{\rm xx}} & = & 
{\displaystyle{-\tilde{f}_{N}^{2}+\tilde{f}_{N}^{\prime 2}-4 
\overwithdelims.. \tilde{f}_{N}^{2}-\tilde{f}_{N}^{\prime 2}+
4\left( \tilde{f}_{N}^{\prime }-1\right) }}%
\\ 
\tilde{r}_{{\rm xy}} & = & -4i%
{\displaystyle{\tilde{f}_{N} \overwithdelims.. \tilde{f}_{N}^{2}-
\tilde{f}_{N}^{\prime 2}+4\left( \tilde{f}_{N}^{\prime }-1\right) }}%
\end{array}
\right. \ . 
\]

%
%%%%%%%%%%%%%%%%%%%%%%%%%%%%%%%%%%%%%%%%%%%%%%%%%%%%%%%%%%%%%%%%%%%%%%%
%   REFERENCES
%%%%%%%%%%%%%%%%%%%%%%%%%%%%%%%%%%%%%%%%%%%%%%%%%%%%%%%%%%%%%%%%%%%%%%%
%
\bibliographystyle{prsty}

%
%%%%%%%%%%%%%%%%%%%%%%%%%%%%%%%%%%%%%%%%%%%%%%%%%%%%%%%%%%%%%%%%%%%%%%%
%   Table: SOLUTIONS OF THE HELMHOLTZ EQUATION
%%%%%%%%%%%%%%%%%%%%%%%%%%%%%%%%%%%%%%%%%%%%%%%%%%%%%%%%%%%%%%%%%%%%%%%
%
\begin{table}[tbhp]
\centering
\begin{tabular}{|c|c|c|c|c|}
$k$ & $1$ & $2$ & $3$ & $4$ \\ \hline
${\cal E}_{p{\rm x}}^{(k)}$ & arbitrary & $i{\cal E}_{p{\rm y}}^{(2)}$ & 
arbitrary & $i{\cal E}_{p{\rm y}}^{(4)}$ \\ \hline
${\cal E}_{p{\rm y}}^{(k)}$ & $i{\cal E}_{p{\rm x}}^{(1)}$ & arbitrary & $i%
{\cal E}_{p{\rm x}}^{(3)}$ & arbitrary \\ \hline
${\cal E}_{p{\rm z}}^{(k)}$ & 0 & 0 & 0 & 0
\end{tabular}
\caption[Solutions of the Helmholtz equation.]
    {\label{tab:solhelm4polgeomappni} 
      Solutions of the Helmholtz equation (\ref{eq:pwphelm}) for polar
      geometry and normal incidence neglecting the difference in the
      diagonal elements of the layer--resolved
      permittivity. $\protect{{\cal E}^{(k)}_{p\mu}}$ is the amplitude
      of the electric field in layer $\protect{p}$ for beam $\protect{
      k}$.}
\end{table}

%
%%%%%%%%%%%%%%%%%%%%%%%%%%%%%%%%%%%%%%%%%%%%%%%%%%%%%%%%%%%%%%%%%%%%%%%
%   Table: SOLUTIONS OF THE CURL MAXWELL EQUATION
%%%%%%%%%%%%%%%%%%%%%%%%%%%%%%%%%%%%%%%%%%%%%%%%%%%%%%%%%%%%%%%%%%%%%%%
%
\begin{table}[tbph]
\centering
\begin{tabular}{|c|c|c|c|c|}
$k$ & $1$ & $2$ & $3$ & $4$ \\ \hline
${\cal H}_{p{\rm x}}^{(k)}$ & $i{\cal E}_{p{\rm x}}^{(1)}\sqrt{\tilde{%
\epsilon}_{{\rm xx}}^{p}+i\tilde{\epsilon}_{{\rm xy}}^{p}}$ & ${\cal E}%
_{p{\rm y}}^{(2)}\sqrt{\tilde{\epsilon}_{{\rm xx}}^{p}-i\tilde{\epsilon%
}_{{\rm xy}}^{p}}$ & $-i{\cal E}_{p{\rm x}}^{(3)}\sqrt{\tilde{\epsilon}_{%
{\rm xx}}^{p}+i\tilde{\epsilon}_{{\rm xy}}^{p}}$ & $-{\cal E}_{p{\rm y}%
}^{(4)}\sqrt{\tilde{\epsilon}_{{\rm xx}}^{p}-i\tilde{\epsilon}_{{\rm xy%
}}^{p}}$ \\ \hline
${\cal H}_{p{\rm y}}^{(k)}$ & $-{\cal E}_{p{\rm x}}^{(1)}\sqrt{\tilde{%
\epsilon}_{{\rm xx}}^{p}+i\tilde{\epsilon}_{{\rm xy}}^{p}}$ & $-i{\cal %
E}_{p{\rm y}}^{(2)}\sqrt{\tilde{\epsilon}_{{\rm xx}}^{p}-i\tilde{%
\epsilon}_{{\rm xy}}^{p}}$ & ${\cal E}_{p{\rm x}}^{(3)}\sqrt{\tilde{%
\epsilon}_{{\rm xx}}^{p}+i\tilde{\epsilon}_{{\rm xy}}^{p}}$ & $i{\cal E%
}_{p{\rm y}}^{(4)}\sqrt{\tilde{\epsilon}_{{\rm xx}}^{p}-i\tilde{%
\epsilon}_{{\rm xy}}^{p}}$ \\ \hline
${\cal H}_{p{\rm z}}^{(k)}$ & 0 & 0 & 0 & 0
\end{tabular}
\caption[Solutions of the curl Maxwell equation.]
    {\label{tab:solcurl4polgeomappni}
      Solutions of the curl Maxwell equation (\ref{eq:pwpcurl}) for
      polar geometry and normal incidence neglecting the difference in
      the diagonal elements of the layer--resolved permittivity
      $\protect{{\bbox{\epsilon}}^{p}}$. $\protect{{\cal H}_{p\mu
      }^{(k)}}$ is the amplitude of the magnetic field in layer
      $\protect{p}$ for beam $\protect{k}$.}
\end{table}

\vfill
\newpage
%
%%%%%%%%%%%%%%%%%%%%%%%%%%%%%%%%%%%%%%%%%%%%%%%%%%%%%%%%%%%%%%%%%%%%%%%
%   Figure 1.: MACROSCOPIC MODEL OF A LAYER SYSTEM
%%%%%%%%%%%%%%%%%%%%%%%%%%%%%%%%%%%%%%%%%%%%%%%%%%%%%%%%%%%%%%%%%%%%%%%
%
%=======================================================================
% FIGURE: macroscopic model of a layer system
 
%=======================================================================
%
\vfill
\newpage
%
%
%%%%%%%%%%%%%%%%%%%%%%%%%%%%%%%%%%%%%%%%%%%%%%%%%%%%%%%%%%%%%%%%%%%%%%%
%   Figure 2.: FCC(111)-Pt BULK PERMITTIVITY
%%%%%%%%%%%%%%%%%%%%%%%%%%%%%%%%%%%%%%%%%%%%%%%%%%%%%%%%%%%%%%%%%%%%%%%
%
%=======================================================================
% FIGURE: fcc(111)-Pt bulk permittivity
 
%=======================================================================
%
\vfill
\newpage
%
%%%%%%%%%%%%%%%%%%%%%%%%%%%%%%%%%%%%%%%%%%%%%%%%%%%%%%%%%%%%%%%%%%%%%%%
%   Figure 3.: RELATIVE DIFFERENCES IN THE PERMITTIVITIES
%%%%%%%%%%%%%%%%%%%%%%%%%%%%%%%%%%%%%%%%%%%%%%%%%%%%%%%%%%%%%%%%%%%%%%%
%
%
%=======================================================================
% FIGURE: differences in the layer--resolved permittivities
 
%=======================================================================
%
%%%%%%%%%%%%%%%%%%%%%%%%%%%%%%%%%%%%%%%%%%%%%%%%%%%%%%%%%%%%%%%%%%%%%%%
%   Figure 4.: PMOKE 2X2 MATRIX TECHNIQUE 
%%%%%%%%%%%%%%%%%%%%%%%%%%%%%%%%%%%%%%%%%%%%%%%%%%%%%%%%%%%%%%%%%%%%%%%
%
%=======================================================================
% FIGURE: PMOKE 2x2 matrix technique 
 
%=======================================================================
%
\vfill
\newpage
%
%%%%%%%%%%%%%%%%%%%%%%%%%%%%%%%%%%%%%%%%%%%%%%%%%%%%%%%%%%%%%%%%%%%%%%%
%   Figure 5.: PMOKE TWO-MEDIA APPROACH
%%%%%%%%%%%%%%%%%%%%%%%%%%%%%%%%%%%%%%%%%%%%%%%%%%%%%%%%%%%%%%%%%%%%%%%
%
%=======================================================================
% FIGURE: PMOKE two-media approach
 
%=======================================================================
%

\end{document}